\documentclass[11pt,a4paper]{article}
\usepackage[margin=25mm]{geometry}
\usepackage[T1]{fontenc}
\usepackage[utf8]{inputenc}
\usepackage{newtxtext}
\usepackage{amsmath,amssymb,amsfonts}
\usepackage{newtxmath}
\usepackage{graphicx,bm,textgreek,enumitem,makecell,tikz}
\usepackage[normalem]{ulem}
\usepackage[dvipsnames]{xcolor}
\usepackage{url,soul,comment,appendix}
\usepackage[numbers,sort&compress]{natbib}
\usepackage[font=small,labelfont=bf,labelsep=period]{caption}
\usepackage{setspace,titlesec,etoolbox}
\usepackage[hidelinks]{hyperref}
\titleformat{\section}{\large\bfseries}{\thesection.}{0.6em}{}
\titleformat{\subsection}{\normalsize\bfseries}{\thesubsection.}{0.6em}{}

\graphicspath{{figures/}}
\newcommand{\R}[1]{#1}

\sethlcolor{yellow}
\soulregister\ref7
\soulregister\eqref7
\soulregister\cite7
\soulregister\mbox7
\soulregister\textsubscript7
\soulregister\textsuperscript7
\ifdefined\ShowChanges
  \newcommand{\Rev}[1]{\hl{#1}}
  \newcommand{\RevMath}[1]{\colorbox{yellow}{$\displaystyle #1$}}
\else
  \newcommand{\Rev}[1]{#1}
  \newcommand{\RevMath}[1]{#1}
\fi
\begin{document}
{\raggedright\LARGE\bfseries Emergence of moir\'{e} magnetic chaos in twisted bilayer CrI\textsubscript{3}\par}

\vspace{12pt}
{\setlength{\parindent}{0pt}
Gyuyoung Park, OukJae Lee, Kyoung-Min Kim*

\vspace{8pt}
Gyuyoung Park, OukJae Lee\\
Center for Semiconductor Technology, Korea Institute of Science and Technology, Seoul, Republic of Korea

\vspace{5pt}
Kyoung-Min Kim\\
Asia Pacific Center for Theoretical Physics, Pohang, Gyeongbuk, 37673, Republic of Korea\\
Department of Physics, Pohang University of Science and Technology, Pohang, Gyeongbuk 37673, Republic of Korea

*\href{mailto:kyoungmin.kim@apctp.org}{kyoungmin.kim@apctp.org}
}

\begin{center}\small (Dated: August 3, 2026)\end{center}
\begin{abstract}
The study of magnetic chaos has traditionally focused on macroscopic variables under external driving. Here we demonstrate a new type of magnetic chaos, termed moir\'{e} magnetic chaos, associated with mesoscopic magnetic domain variables in twisted bilayer CrI\textsubscript{3} without external driving. The domains are stabilized by a characteristic interlayer exchange frustration, which supplies the multiple dynamical degrees of freedom required for \Rev{transient chaos}. Through micromagnetic simulations, we show that relaxation toward moir\'{e} magnetic textures is extremely sensitive to minute \Rev{spatially distributed} perturbations of the initial state, characterized by \Rev{positive} finite-time Lyapunov exponents and a final-state sensitivity that persists over \Rev{three} decades of perturbation amplitude. Statistical analysis further reveals that the resulting domain configurations are stochastic and pairwise uncorrelated. Our results identify a form of microscopic, undriven chaos in twisted magnets that extends nonlinear magnetism beyond the conventional driven regime.
\end{abstract}

\section{Introduction}

The study of magnetic chaos has uncovered intriguing nonlinear dynamics that extend well beyond the linear regime. In ferromagnets, the uniform magnetization, represented by a normalized three-dimensional vector, can be driven by external forces into chaotic behavior with extreme sensitivity to initial conditions \cite{PhysRevB.74.054417, PhysRevLett.99.134101, Montoya2019, PhysRevB.100.224422, TANIGUCHI2019281, PhysRevB.109.214412, Taniguchi2024}. Similar chaotic dynamics have been observed in nonuniform topological spin textures such as magnetic vortices and skyrmions, particularly in their core positions or topological charges \cite{Petit-Watelot2012, PhysRevB.88.014432, PhysRevLett.123.147701, PhysRevB.99.054402, PhysRevResearch.3.043216, PhysRevB.108.174441, PhysRevB.109.014422, PhysRevB.109.174420}. Such macroscopic variables describe a global magnetic state that inherently possesses fewer than three dynamical degrees of freedom (DOF). The application of external forces is therefore essential to raise the system's dimensionality to three or more, satisfying the requirement of the Poincar\'{e}--Bendixson theorem for autonomous chaos \cite{strogatz2001nonlinear, alligood1997chaos, ott2002chaos}. High-frequency or high-amplitude external drives are typically employed for this purpose, including spin-transfer torque \cite{PhysRevB.74.054417, PhysRevLett.99.134101, Montoya2019, TANIGUCHI2019281, PhysRevB.100.224422, PhysRevLett.123.147701}, time-dependent fields \cite{PhysRevB.108.174441, PhysRevB.99.054402, PhysRevB.88.014432}, perpendicular magnetic anisotropy \cite{Taniguchi2024, PhysRevB.109.174420}, and feedback effects \cite{PhysRevB.109.214412, PhysRevResearch.3.043216}. Only a limited number of studies have highlighted intrinsic mechanisms, such as periodic core reversal of vortices (vortex chaos) \cite{Petit-Watelot2012, Park2025} and strong nonlinearity of the magnetization dynamics in a ferrimagnet \cite{PhysRevB.109.014422}. Given the intrinsic nonlinearity of complex magnetic interactions, however, chaos could plausibly emerge in non-macroscopic variables, or even without any external driving. The two possibilities represent a significant gap that warrants deeper investigation in nonlinear magnetism.

Of particular interest here are moir\'{e} magnetic textures in twisted van der Waals (vdW) magnets. Twisted vdW magnets are artificially fabricated heterostructures in which two magnetic layers are rotated relative to one another by a designated twist angle. A prototypical system is the twisted multilayer of chromium triiodide (CrI\textsubscript{3}), investigated intensively both theoretically \cite{PhysRevResearch.3.013027, doi:10.1073/pnas.2000347117, Akram2021, Ghader2022, Zheng2022, Kim2023, PhysRevB.108.L100401, PhysRevB.108.174440, Ganguli2023, https://doi.org/10.1002/admi.202300188, Kim2024, Lee_2024} and experimentally \cite{doi:10.1126/science.abj7478, Xie2022, Xu2022, Xie2023, Cheng2023, Li2024, Yang2024, Wong2026}. In twisted bilayer CrI\textsubscript{3}, the rotation produces a spatial alternation of the interlayer exchange between ferromagnetic (FM) and antiferromagnetic (AFM) coupling, governed by the local stacking symmetry [Fig.~\ref{fig:model}a]. The resulting frustration induces moir\'{e} magnetic textures, characterized by elliptic nanoscale magnetic domains arranged periodically with the moir\'{e} superlattice period \cite{doi:10.1073/pnas.2000347117, PhysRevResearch.3.013027}. Within each AFM patch, the magnetic moments of adjacent layers align antiparallel, stabilized by a strong out-of-plane anisotropy, whereas parallel alignment persists in the surrounding FM background. Each AFM patch possesses two degenerate configurations. In one, the top-layer moments point in a given out-of-plane direction and the bottom-layer moments point oppositely; the other is its time reversal \cite{Kim2023}. We refer to the binary local degree of freedom of each patch as its ``polarization.'' Atomistic simulations show that a combinatorially large number of metastable configurations emerge because each patch selects one of the two polarizations independently \cite{PhysRevB.108.L100401}. The configurations are nearly degenerate, owing to the symmetry between the top and bottom layers and the weak coupling between adjacent AFM patches. Twisted bilayer CrI\textsubscript{3} therefore offers an attractive platform for complex magnetic behavior governed by a periodic array of AFM patches with a high level of metastability.

In this paper we demonstrate that twisted CrI\textsubscript{3} bilayers can display \Rev{transient} magnetic chaos even without external forces, enabled by their interlayer exchange frustration. The frustration prevents the magnetic moments from acting collectively and instead makes them independent dynamical DOFs, so that the system inherently meets the minimum requirement of three DOFs for chaos. Through micromagnetic simulations of twisted bilayer CrI\textsubscript{3}, we show that relaxation to the moir\'{e} magnetic textures is strongly sensitive to the initial conditions. A minute \Rev{spatially distributed} perturbation of the magnetization drives a large divergence in the polarizations of multiple AFM patches, producing substantial positive \Rev{finite-time} Lyapunov exponents \Rev{during} \Rev{relaxation} (Sec.~\ref{sec:damage_spreading}). We term the extreme sensitivity to initial conditions ``moir\'{e} magnetic chaos.'' We further show that the per-patch Shannon entropy reaches 99.\Rev{99}\% of its maximum and that the pairwise correlations between polarizations are statistically indistinguishable from independence, confirming that the polarizations are stochastic and pairwise uncorrelated (Sec.~\ref{sec:stat_anal}). Together, the findings support the classification of twisted vdW magnets as a new class of high-dimensional chaotic systems.

\section{Results}

\subsection{Frustration-induced nonlinearity and high metastability.} \label{sec:model}

\begin{figure*}[t!]
    \centering
    \includegraphics[width=\textwidth]{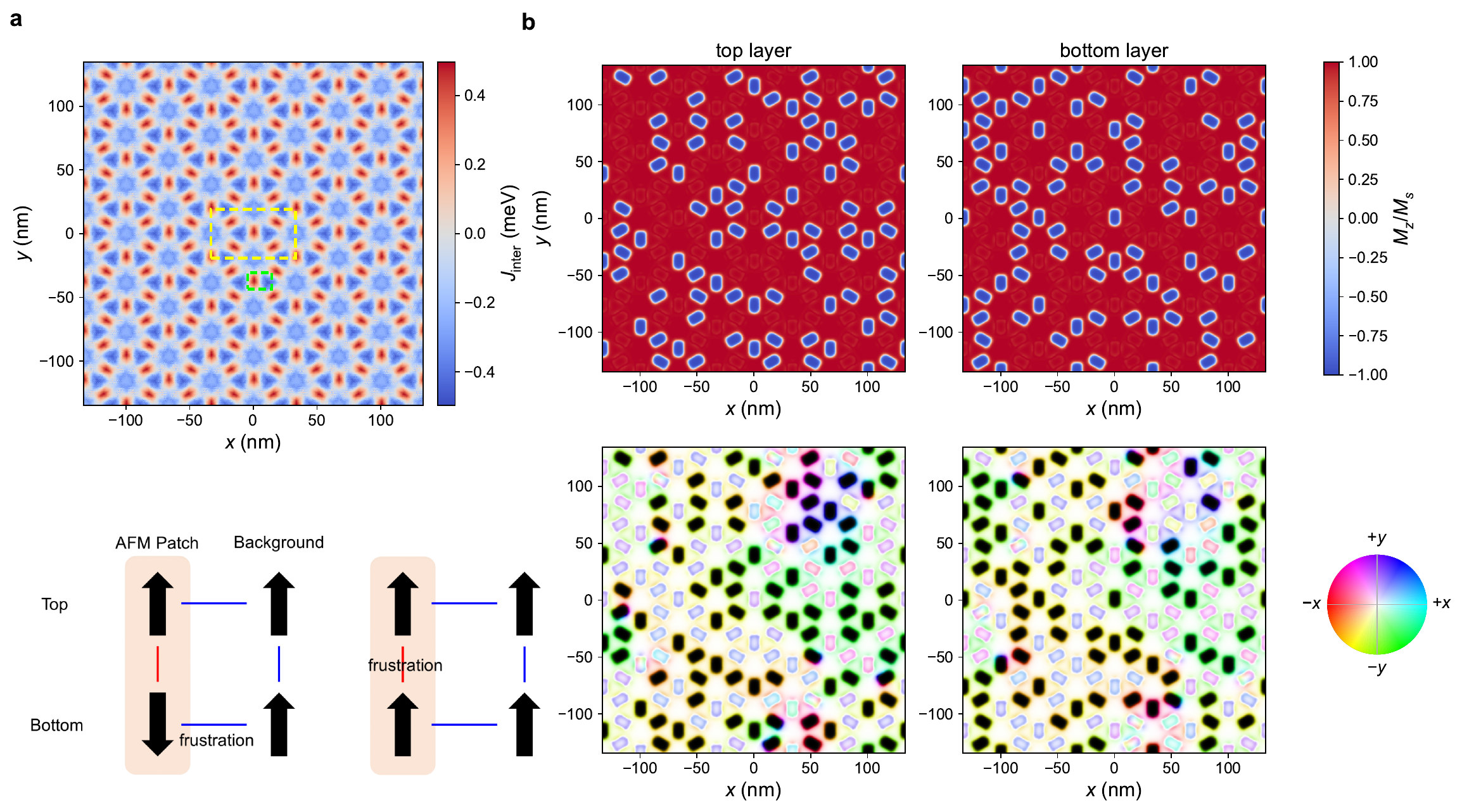}
    \caption{
    \textbf{Moir\'{e} magnetic textures in twisted bilayer CrI\textsubscript{3}.} \textbf{a}, Interlayer exchange coupling map $J_{\mathrm{inter}}(\bm{r})$ used in the simulations, tiled across the moir\'{e} superlattice. Blue (red) marks ferromagnetic (antiferromagnetic) coupling, $J_{\mathrm{inter}} < 0$ ($J_{\mathrm{inter}} > 0$). The yellow dashed box outlines one moir\'{e} unit cell at the centre of the map; the small green dashed box just below it isolates a single AFM patch (red) next to the surrounding FM background (blue). \textbf{b}, A representative metastable configuration obtained by relaxing a random initial state (a uniformly out-of-plane-polarised background carrying random in-plane fluctuations; see Methods). Top row, out-of-plane magnetization $M_z / M_s$ of the top (left) and bottom (right) layers. The ferromagnetic background ($M_z/M_s \approx +1$, red) hosts reversed domains ($M_z/M_s \approx -1$, blue) that appear at complementary AFM patches in the two layers, as favoured by the local AFM coupling. The reversed domains are of two kinds, skyrmion-like and bubble-like, and their detailed classification is given in Fig.~\ref{fig:skbubble}. Bottom row, in-plane magnetization of the two layers, encoded as the in-plane angle $\mathrm{atan2}(M_y, M_x)$ through the hue--saturation--value (HSV) colour wheel (inset). Below \textbf{a}, schematic of the interlayer frustration at a single AFM patch, showing its top and bottom layers and the surrounding ferromagnetic background. In the antiparallel (AFM) state the interlayer AFM bond (red) is satisfied but the bottom-layer bond (blue) to the background is frustrated; in the parallel (FM) state the bonds to the background are satisfied but the interlayer AFM bond is frustrated. Neither state satisfies every bond (labelled ``frustration''), which is the origin of the local nonlinearity.
    }
    \label{fig:model}
\end{figure*}

We consider a twisted bilayer of CrI\textsubscript{3} in which two honeycomb-lattice monolayers are stacked with a small twist angle \Rev{\mbox{$\theta = 1.023^\circ$}}. The slowly varying local stacking makes the interlayer exchange alternate between ferromagnetic and antiferromagnetic regions across the moir\'{e} cell. The magnetic energy is described by the \Rev{classical} Heisenberg spin model:
\begin{equation}
\begin{aligned}
    \label{eq:spinH}
    \RevMath{H} = & -J_{\mathrm{intra}}\sum_{l=t,b} \sum_{\langle i, j\rangle} \bm{S}_{i}^{l} \cdot \bm{S}_{j}^{l} - A \sum_{l=t,b} \sum_{i} \big(\bm{S}_{i}^{l}\cdot\hat{\bm{z}}\big)^2 + \sum_{i, j} J_{\mathrm{inter},ij} \bm{S}_{i}^{t} \cdot \bm{S}_{j}^{b} \\
    & -\frac{\mu_0(g_s\mu_B)^2}{4\pi} \sum_{l,l'} \sum_{i\neq j} \frac{1}{|\bm{r}_{ij}|^3}\big[3(\bm{S}_i^l\cdot\hat{\bm{r}}_{ij})(\bm{S}_j^{l'}\cdot\hat{\bm{r}}_{ij})-\bm{S}_i^{l}\cdot\bm{S}_j^{l'}\big],
\end{aligned}
\end{equation}
where $\bm{S}_{i}^{l}$ is the unit-magnitude spin at site $i$ on layer $l \in \{t, b\}$ (top or bottom), the inner sum $\langle i, j\rangle$ is restricted to nearest-neighbour pairs within a layer, $J_{\mathrm{intra}} > 0$ is the intralayer FM exchange, $A > 0$ is the single-ion anisotropy along the out-of-plane axis $\hat{\bm{z}}$, and $J_{\mathrm{inter},ij}$ is the stacking-dependent interlayer exchange coupling between top-layer site $i$ and bottom-layer site $j$. The map $J_{\mathrm{inter}}(\bm{r})$ for this \Rev{\mbox{$\theta = 1.023^\circ$}} geometry is obtained from first principles using density-functional theory combined with the magnetic force theorem\cite{Kim2023SpinHam} [Fig.~\ref{fig:model}a]. The last term represents dipole--dipole interactions, where $\mu_0$ is the vacuum permeability, $g_s \approx 2$ is the electron spin $g$-factor, $\mu_B$ is the Bohr magneton, $\bm{r}_{ij}$ is the displacement vector from site $i$ to site $j$ (with $\hat{\bm{r}}_{ij} = \bm{r}_{ij} / |\bm{r}_{ij}|$), and the layer indices $l, l'$ run over $\{t, b\}$. We employ $J_{\mathrm{intra}} = 2.12$~meV and $A = 0.087$~meV for CrI\textsubscript{3}~\cite{Kim2023SpinHam}, within the range studied for the chromium-trihalide family of two-dimensional Heisenberg ferromagnets~\cite{LadoFR2017}. Quantitative outcomes reported below, such as the \Rev{local} \Rev{growth rates} and the \Rev{thermal} \Rev{response}, scale with these parameters, whereas the statistical conclusions concern the structure of the ensemble.

The interlayer coupling map $J_{\mathrm{inter}}(\bm{r})$ in Fig.~\ref{fig:model}a comprises a dominant FM background ($J_{\mathrm{inter}} < 0$, blue) punctuated by isolated AFM patches ($J_{\mathrm{inter}} > 0$, red) on the moir\'{e} lattice. Each AFM patch is frustrated (schematic below Fig.~\ref{fig:model}a). Its AFM interlayer bond favours antiparallel top/bottom moments and admits two degenerate polarizations (top-up/bottom-down and its time reversal), while the surrounding FM bonds to the common background oppose the local order. The competition between a patch and its FM environment makes the local relaxation dynamics nonlinear. Because neighbouring patches are only weakly coupled, each patch selects its polarization quasi-independently. The outcome is an exponentially large manifold of nearly degenerate metastable states, $2^{N}$ for $N$ AFM patches, which is the high metastability that underlies the chaotic behaviour analysed below.

We confirm this near-degeneracy directly from the \Rev{original \mbox{$1000$}-run} ensemble. Re-evaluating the total energy of the relaxed configurations, we find that the \Rev{sampled} \Rev{ensemble} spans only $E_{\max} - E_{\min} \approx 48$~meV for the full bilayer, a relative spread of $\Delta E / |E| \approx 1.8 \times 10^{-4}$, or roughly $0.3$~meV per AFM patch (Fig.~\ref{fig:dos}). The density of states is accurately Gaussian, with a standard deviation of $\sigma \approx 7.5$~meV, as expected by the central limit theorem when the total energy is a sum of many nearly independent per-patch contributions. 

A representative metastable texture is shown in Fig.~\ref{fig:model}b, obtained by relaxing one random initial state (the out-of-plane-polarised, domain-free background dressed with random in-plane fluctuations; see Methods) and consistent with previous observations\cite{PhysRevB.108.174440, PhysRevResearch.3.013027, Kim2023, PhysRevB.103.L140406, Kim2024}. The top row shows the out-of-plane magnetization $m_z$ of the two layers. Both share the FM background ($m_z \approx +1$, red), and at the AFM patches reversed domains ($m_z \approx -1$, blue) nucleate in one layer while the other retains the background orientation, so the two layers host complementary, non-overlapping domains. Strikingly, the reversed domains occupy a seemingly random subset of the AFM patches and form a chaotic-looking pattern that differs from run to run. Which patches reverse is set entirely by the initial condition, since distinct random seeds relax to distinct arrangements from the same energy landscape, a direct manifestation of the high metastability. The bottom row shows the corresponding in-plane magnetization (encoded through the HSV colour wheel), which reveals that the reversed domains are of two kinds, skyrmion-like and bubble-like. Excluding domains that touch the simulation boundary, the moir\'{e} cell hosts $N = 157$ AFM patches, whose classification and numbering scheme are given in the Supplementary Information.

\subsection{Damage spreading and \Rev{transient} chaos.}\label{sec:damage_spreading}

To quantify the chaotic nature of the magnetization dynamics, we perform a damage-spreading analysis probing sensitivity to \Rev{small} perturbations (see Methods). We prepare a reference trajectory by relaxing the system for $t = 2$~ns at zero temperature, then generate perturbed trajectories by \Rev{applying} a \Rev{spatially} \Rev{distributed} \Rev{tangent} \Rev{perturbation} \Rev{of amplitude} $\varepsilon$. Sweeping $\varepsilon$ over \Rev{three} decades from \Rev{\mbox{$10^{-5}$}} to \Rev{\mbox{$10^{-2}$}}, we compute the root-mean-square (RMS) magnetization difference:
\begin{equation}
    \delta m(t) = \sqrt{\langle |\bm{m}_\text{pert}(\bm{r}, t) - \bm{m}_\text{ref}(\bm{r}, t)|^2 \rangle},
\end{equation}
where $\bm{m}_\text{ref}(\bm{r}, t)$ and $\bm{m}_\text{pert}(\bm{r}, t)$ are the local magnetization vectors of the reference and perturbed trajectories, respectively, and $\langle \cdot \rangle$ denotes the spatial average over \Rev{both layers of} the simulation cell. We define the finite-amplitude \Rev{amplification} \Rev{rate} as
\begin{equation}
\RevMath{\Lambda(t,\varepsilon) = \frac{1}{t} \ln \frac{\delta m(t)}{\varepsilon}.}
    \label{eq:lyapunov}
\end{equation}

\begin{figure*}[t!]
    \centering
    \includegraphics[width=\textwidth]{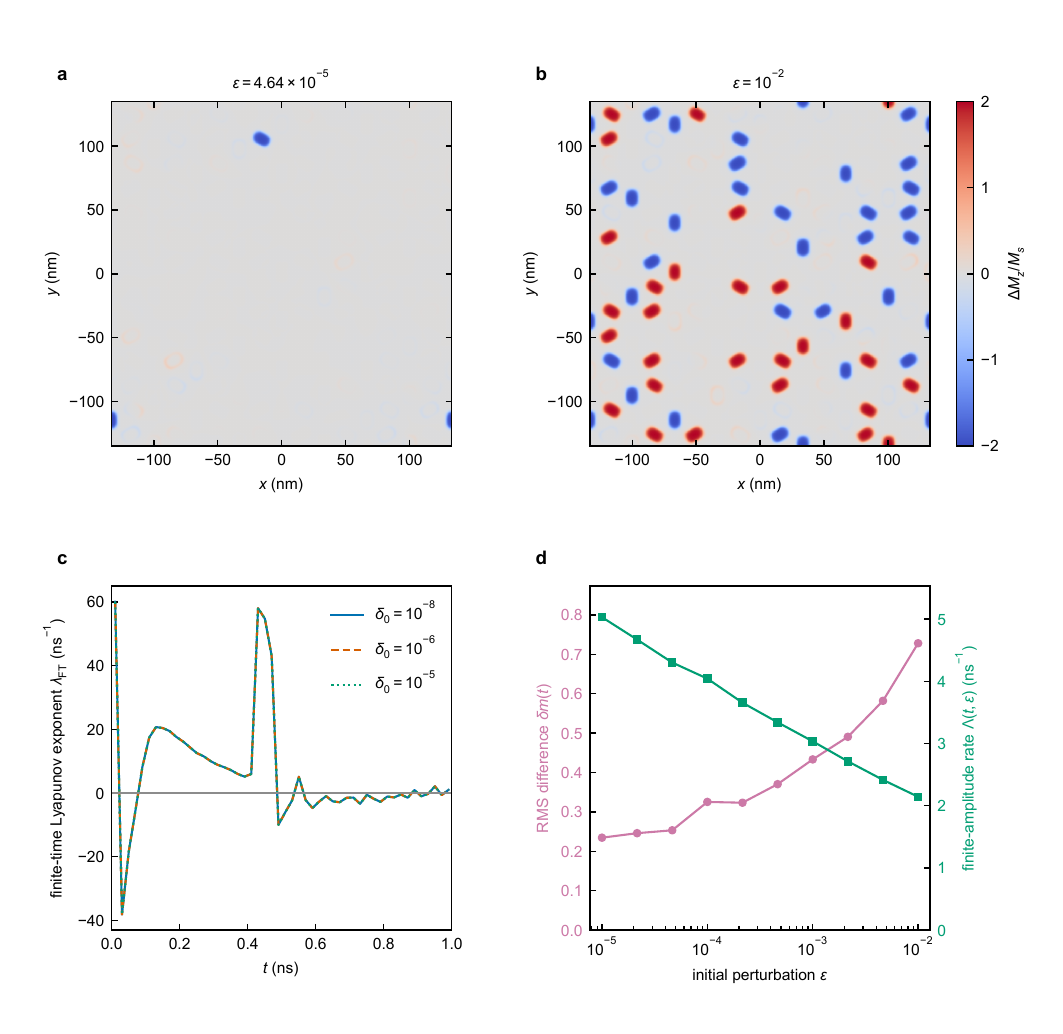}
    \caption{
    \textbf{Damage spreading and \Rev{transient} chaos.} \textbf{a}, Spatial map of the signed top-layer magnetization difference \Rev{\mbox{$\Delta M_z(\bm{r})/M_s$}} between a perturbed and the reference trajectory at $t = 2$~ns, for the perturbation \Rev{\mbox{$\varepsilon = 4.64\times10^{-5}$}}. Red and blue mark domains that \Rev{differ} relative to the reference; white marks the remaining unchanged region (Hamming fraction \Rev{\mbox{$H/N = 1/157$}}, where \Rev{\mbox{$N = 157$}} is the number of \Rev{interior} \Rev{AFM patches}). \textbf{b}, The same map for the largest perturbation \Rev{\mbox{$\varepsilon = 10^{-2}$}} (\Rev{\mbox{$H/N = 56/157$}}). \textbf{c}, \Rev{Finite-time} Lyapunov \Rev{exponents} \Rev{for \mbox{$\delta_0=10^{-8}$}, \mbox{$10^{-6}$} and \mbox{$10^{-5}$} (blue solid, orange dashed and green dotted), averaged over \mbox{$20$}~ps.} \textbf{\Rev{d}}\Rev{, Finite-amplitude amplification rate \mbox{$\Lambda(t,\varepsilon)$} (green squares, right) and RMS separation \mbox{$\delta m$} (pink circles, left)} at \Rev{\mbox{$t=2$}}~ns \Rev{versus} $\varepsilon$\Rev{;} \Rev{lines} \Rev{connect} the \Rev{ten} \Rev{measurements.} \Rev{Panels} \Rev{a,} \Rev{b} \Rev{and} \Rev{d} \Rev{share} \Rev{one initial state and perturbation direction; c uses another tangent direction and double precision with \mbox{$6.25$}~fs steps}.
    }
    \label{fig:lyapunov}
\end{figure*}

Figure~\ref{fig:lyapunov}a,b maps the \Rev{top-layer} magnetization difference \Rev{\mbox{$\Delta M_z(\bm r)/M_s$}} between a perturbed and the reference trajectory at $t = 2$~ns. \Rev{For} \Rev{\mbox{$\varepsilon = 4.64\times10^{-5}$}}, \Rev{one} \Rev{patch} \Rev{has} reversed \Rev{its} binary state, giving a Hamming fraction \Rev{\mbox{$H/N=1/157$}}, where $H$ counts the \Rev{patches} whose state differs between the perturbed and reference trajectories and \Rev{\mbox{$N=157$}} is the total number of \Rev{interior} AFM patches. Raising $\varepsilon$ to \Rev{\mbox{$10^{-2}$}} \Rev{increases} the fraction \Rev{to \mbox{$H/N=56/157$}.}

\Rev{At \mbox{$t=2$}~ns, the finite-amplitude amplification rate (Fig.~\ref{fig:lyapunov}d) decreases monotonically from \mbox{$\Lambda(t,\varepsilon)\approx5.03$}~ns\mbox{$^{-1}$}} at \Rev{\mbox{$\varepsilon=10^{-5}$}} to \Rev{\mbox{$\Lambda(t,\varepsilon)\approx2.14$}~ns\mbox{$^{-1}$}} \Rev{at} \Rev{\mbox{$\varepsilon=10^{-2}$}}.

The \Rev{finite-time} Lyapunov exponent \Rev{is measured by Benettin rescaling every \mbox{$\tau=1$}~ps:}
\begin{equation}
\RevMath{\lambda_{\mathrm{FT}}(t_k;\tau)=\tau^{-1}\ln(d_k/\delta_0),}
\label{eq:local_growth}
\end{equation}
\Rev{where \mbox{$d_k$} is the per-spin RMS separation before rescaling. Figure~\ref{fig:lyapunov}c shows \mbox{$20$}~ps block averages. Positive finite-time Lyapunov exponents followed by late contraction} (Fig.~\ref{fig:lyapunov}c) \Rev{supports} a \Rev{transient-chaos} \Rev{interpretation} of the \Rev{relaxation;} \Rev{numerical} \Rev{controls} \Rev{are} \Rev{given} \Rev{in Supplementary Sec.~\ref{sec:lyapunov}.}

\Rev{Higher damping suppresses final-domain sensitivity in} the \Rev{tested} \Rev{trajectory} \Rev{pairs} \Rev{(Supplementary} \Rev{Fig.~\ref{fig:damping_sensitivity}).} \Rev{Transient} \Rev{amplification} and \Rev{final-domain} \Rev{sensitivity} \Rev{persist} across the \Rev{tested} \Rev{twist} \Rev{angles} \Rev{for} \Rev{\mbox{$\varepsilon=10^{-3}$}} (\Rev{Supplementary} Fig.~\Rev{\ref{fig:twist_sensitivity}}).

\subsection{Statistical characterization.} \label{sec:stat_anal}

Section~\ref{sec:damage_spreading} established sensitivity to \Rev{small} perturbations of a single reference trajectory. We now characterize the ensemble of \Rev{metastable configurations} accessed by distinct random initial conditions. From \Rev{10000} independent relaxation runs (see Methods), each terminal configuration is recorded as a binary string over the $N = 157$ enumerated AFM patches, with a patch set to $-1$ if it hosts a reversed domain ($m_z \approx -1$) and $+1$ otherwise. The local configuration is in fact ternary (a reversed domain in the top layer, in the bottom layer, or none); the binary bit records the sign of the top-layer $m_z$, and hence the top-layer occupation.

\begin{figure*}[t!]
    \centering
    \includegraphics[width=\textwidth]{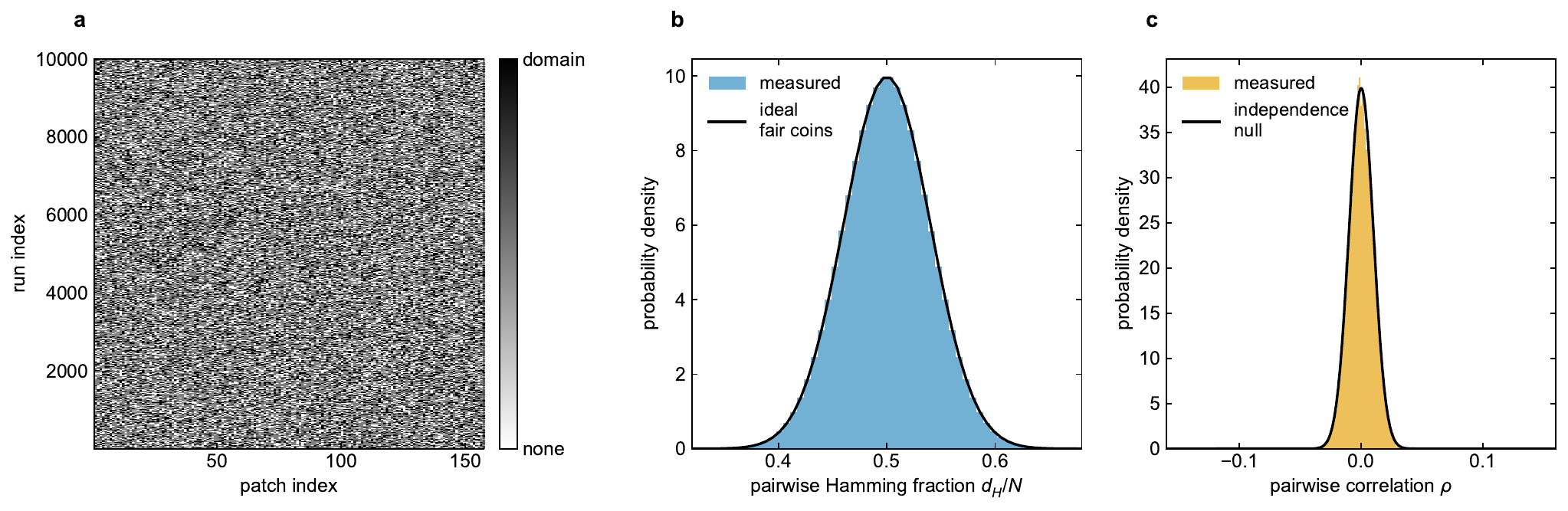}
    \caption{
    \textbf{Random, uncorrelated bits from deterministic dynamics.} \textbf{a}, Raw ensemble raster. Each of the \Rev{10000} rows is the $157$-bit string produced by one deterministic relaxation from a random initial state (black, a reversed domain forms at that AFM patch; white, none). The raster is visually indistinguishable from random noise. \textbf{b}, Distribution of the pairwise Hamming fraction $d_H/N$ over all pairs of runs (bars) versus that of $N$ ideal fair coins, $\mathrm{Bin}(N,1/2)$ (line). The two coincide (measured mean and standard deviation $0.500$ and $0.040$ versus ideal $0.500$ and $0.040$), so any two runs differ in half of their patches. \textbf{c}, Distribution of the Pearson correlation coefficient $\rho$ over all $12{,}246$ patch pairs (bars) versus the finite-sample independence null $\mathcal{N}(0, 1/n_\text{runs})$ (line). The measured width (\Rev{\mbox{$0.010$}}) matches the null (\Rev{\mbox{$1/\sqrt{10000} = 0.010$}}), and no pair survives Bonferroni-corrected significance testing. The enumeration (numbering) of the $N = 157$ AFM patches is given in Fig.~\ref{fig:patch_index}.
    }
    \label{fig:correlation}
\end{figure*}

The raw output of the ensemble is displayed as a raster in Fig.~\ref{fig:correlation}a. Each of the \Rev{10000} rows is the $157$-bit string from one deterministic relaxation, and the image is visually indistinguishable from random noise. Every AFM patch reverses with probability $p_{-1} \approx 0.50$ (mean \Rev{\mbox{$\langle p_{-1} \rangle = 0.4999$}}), so its binary Shannon entropy, with $p_n \equiv p_{-1}^{(n)}$,
\begin{equation}
    H_n = -p_n \log p_n - (1 - p_n) \log(1 - p_n),
    \label{eq:shannon}
\end{equation}
clusters at \Rev{\mbox{$\langle H_n \rangle = 0.6931$}}, which is 99.\Rev{99}\% of the maximum $\log 2 \approx 0.693$~nats (natural units of information), identifying each patch as an unbiased binary variable.

To test whether the $157$-bit strings are drawn uniformly and independently, we compare the Hamming distance between every pair of runs with the ideal fair-coin reference (Fig.~\ref{fig:correlation}b). The measured distribution of the Hamming fraction $d_H/N$ coincides with the binomial $\mathrm{Bin}(N,1/2)$ expected for $N$ independent fair coins, with mean $0.500$ and standard deviation $0.040$ that match the ideal value to three decimals. Any two independent draws thus differ in half of their patches, exactly as for a perfect random-bit source.

We probe inter-patch correlations directly through the Pearson correlation coefficient $\rho_{ij}$ of every patch pair over the \Rev{10000} runs (Fig.~\ref{fig:correlation}c). The distribution of the $N(N-1)/2 = 12{,}246$ coefficients is centred at zero (mean \Rev{\mbox{$-0.00003$}}) with a standard deviation of \Rev{\mbox{$0.010$}}, indistinguishable from the finite-sample independence null $\mathcal{N}(0, 1/n_\text{runs})$ of width \Rev{\mbox{$1/\sqrt{10000} = 0.010$}}, and the largest coefficient is only \Rev{\mbox{$|\rho| = 0.038$}}. After Bonferroni correction for the $12{,}246$ comparisons, not a single patch pair shows a statistically significant correlation. The patch reversals therefore carry no detectable pairwise correlation, \Rev{below} \Rev{the} \Rev{\mbox{$|\rho|=0.046$} detection threshold} at \Rev{\mbox{$n=10000$}}, consistent with statistical independence. We test pairwise correlations here; higher-order dependence is not excluded. A complementary per-patch and pairwise correlation-entropy analysis \Rev{of the original \mbox{$1000$}-run ensemble} is given in Sec.~\ref{sec:entropy} and Fig.~\ref{fig:entropy}.

Taken together, the near-maximal single-patch entropy, the ideal pairwise Hamming statistics, and the pairwise correlations consistent with independence show that the sampled ensemble is consistent with $N = 157$ unbiased, pairwise-uncorrelated top-layer occupancy variables produced by deterministic \Rev{Landau--Lifshitz--Gilbert (}LLG\Rev{)} dynamics alone. We record only the binary top-layer occupancy of each patch, whereas the local state is in fact ternary (a reversed domain in the top layer, in the bottom layer, or neither); a full ternary characterization and a direct demonstration that all $2^{N}$ configurations are dynamically accessible are left to future work, so we refrain from asserting exhaustive access to the $2^{N}$-element manifold here.

\subsection{Thermal robustness of the moir\'{e} domain pattern.} \label{sec:thermal_locking}

A natural concern for the experimental relevance of moir\'{e} magnetic chaos is whether the chaotically selected domain pattern survives thermal fluctuations. To test this robustness, we extend the protocol of Sec.~\ref{sec:model} by replacing the deterministic LLG solver with the stochastic LLG solver (Langevin field obeying the fluctuation--dissipation theorem) and \Rev{evolving} a single \Rev{relaxed} configuration \Rev{separately at} $T = 0$--$29$~K in $1$~K steps \Rev{for \mbox{$2$}~ns}. To quantify how faithfully the moir\'{e} domain pattern is preserved, we define a scalar domain-order parameter from the time-averaged magnetization. For each cell $\bm{r}$ we form the time average of the top-layer out-of-plane magnetization over the \Rev{observation} window $t \in [t_1, t_2]$ (here $t_1 = 1$ and $t_2 = 2$~ns),
\begin{equation}
\overline{m}_z(\bm{r}) = \frac{1}{N_t}\sum_{k=1}^{N_t} m_z^{\rm top}(\bm{r}, t_k), \qquad t_k \in [t_1, t_2],
\label{eq:timeavg}
\end{equation}
and define the domain order as the spatial standard deviation of this time-averaged map,
\begin{equation}
\mathcal{O}(T) \equiv \mathrm{std}_{\bm{r}}\!\left[\overline{m}_z\right]
= \left[\frac{1}{N_{\bm{r}}}\sum_{\bm{r}} \left(\overline{m}_z(\bm{r}) - \overline{m}_z^{\,\rm sp}\right)^{2}\right]^{1/2},
\qquad
\overline{m}_z^{\,\rm sp} = \frac{1}{N_{\bm{r}}}\sum_{\bm{r}} \overline{m}_z(\bm{r}),
\label{eq:order}
\end{equation}
where $N_t$ is the number of \Rev{averaged} frames, $N_{\bm{r}}$ the number of cells, and $\overline{m}_z^{\,\rm sp}$ the spatial mean. Physically, $\mathcal{O}$ is large when persistent reversed domains ($\overline{m}_z \approx -1$) stand out against the ferromagnetic background ($\overline{m}_z \approx +1$), and approaches zero once thermal fluctuations homogenize the time-averaged texture. The Curie temperature of monolayer CrI\textsubscript{3} is $T_C \approx 45$~K~\cite{Huang2017}.

\begin{figure*}[t!]
    \centering
    \includegraphics[width=\textwidth]{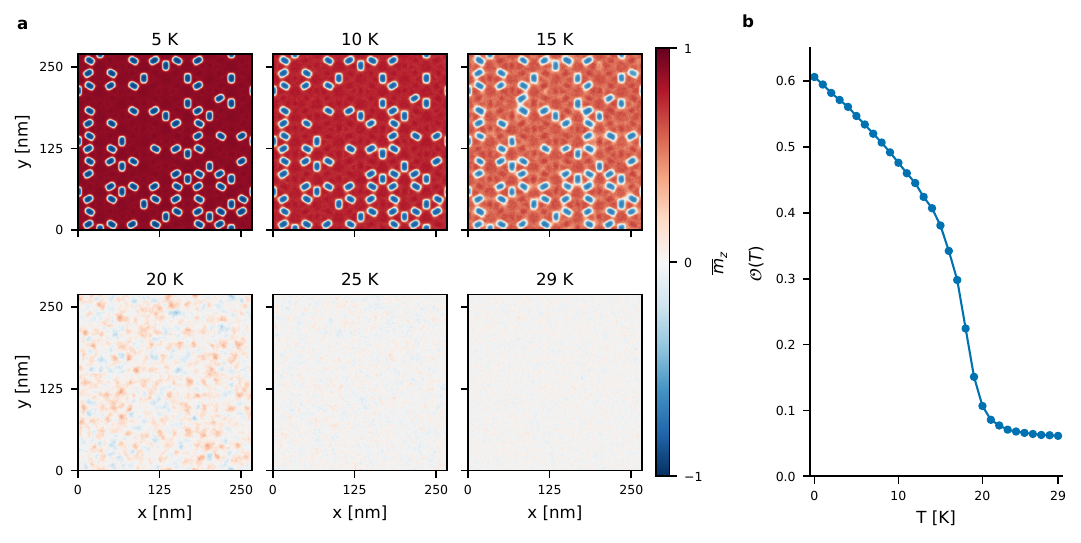}
    \caption{
    \textbf{Thermal robustness of the chaotically selected moir\'{e} domain pattern.} \textbf{a}, Time-averaged top-layer magnetization $\overline{M}_z/M_s$ [Eq.~\eqref{eq:timeavg}], averaged over the \Rev{observation} window $t \in [1,2]$~ns at representative temperatures $T = 5$, $10$, $15$, $20$, $25$, and $29$~K, obtained \Rev{from} \Rev{separate} \Rev{\mbox{$2$}~ns} \Rev{trajectories} \Rev{starting} from \Rev{the same relaxed} $T = 0$ \Rev{configuration}. The reversed domains ($\overline{m}_z \approx -1$, blue) \Rev{persist} \Rev{at} \Rev{low} \Rev{temperature} and progressively \Rev{lose} \Rev{contrast} at higher temperature. \textbf{b}, Domain order $\mathcal{O}$ versus temperature. $\mathcal{O}$ \Rev{decreases} \Rev{from} its $T = 0$ value (\Rev{\mbox{${\approx}0.606$}}) \Rev{to} \Rev{\mbox{${\approx}0.476$} at \mbox{$10$}}~K, steepening near \Rev{\mbox{${\sim}17$}}--\Rev{\mbox{$20$}}~K, and falls to \Rev{\mbox{${\approx}0.061$}} by $29$~K.
    }
    \label{fig:thermal_locking}
\end{figure*}

Figure~\ref{fig:thermal_locking}a shows the time-averaged domain maps across the \Rev{fixed-temperature} sweep. Below ${\sim}10$~K the domain order \Rev{decreases} \Rev{from} its zero-temperature value \Rev{\mbox{$\mathcal{O}\approx0.606$} to \mbox{${\approx}0.476$}} (Fig.~\ref{fig:thermal_locking}b). As the temperature is raised further, $\mathcal{O}$ decreases gradually and then steepens near \Rev{\mbox{$T \approx 17$}}--\Rev{\mbox{$20$}}~K, falling to \Rev{\mbox{$\mathcal{O} \approx 0.061$}} by $29$~K, well below the monolayer Curie point, where the time-averaged texture is nearly featureless. A subset of the reversed domains \Rev{in Fig.~\ref{fig:model}b} are skyrmion-like textures ($Q \approx +1$); these support the full set of eigenmodes characteristic of magnetic skyrmions, namely a low-frequency gyrotropic (translational) mode together with breathing and higher-order azimuthal modes, as resolved by picosecond-cadence spectroscopy in the Supplementary Information (Sec.~\ref{sec:suppl_modes} and Fig.~\ref{fig:suppl_modes}).

\Rev{The} domain pattern \Rev{may} \Rev{be} accessible to slow imaging probes such as scanning quantum magnetometry \cite{doi:10.1126/science.abj7478, Wong2026}. \Rev{Ensemble-level} randomness still requires thermal cycling or repeated quenches, since each quench \Rev{selects} a \Rev{metastable} \Rev{configuration}.

\R{\subsection{Persistence under an applied magnetic field.}} \label{sec:field}

\R{Having \Rev{examined} thermal robustness, we next test whether the domain pattern survives an applied magnetic field, the other perturbation a readout experiment must contend with. Starting from a single relaxed configuration, we ramp a uniform field from $0$ to $2$~T \Rev{in \mbox{$0.1$}~T steps}, separately for an in-plane ($\bm{H} \parallel \hat{\bm{x}}$) and an out-of-plane ($\bm{H} \parallel \hat{\bm{z}}$) orientation\Rev{, relaxing the energy to tolerance \mbox{$10^{-9}$} at each step}, and quantify how much of the zero-field pattern is retained through a domain-pattern survival $S(B)$, the normalised spatial correlation between the field-$B$ and zero-field magnetization maps\Rev{, averaged over both layers}:}
\begin{equation}
\RevMath{S(B)=\frac12\sum_{l=t,b}\frac{\sum_{\bm r}\delta M_z^l(\bm r;B)\,\delta M_z^l(\bm r;0)}{\sqrt{\sum_{\bm r}[\delta M_z^l(\bm r;B)]^2}\,\sqrt{\sum_{\bm r}[\delta M_z^l(\bm r;0)]^2}}.}
\label{eq:survival}
\end{equation}
\R{where \Rev{\mbox{$\delta M_z^l(\bm r;B)=M_z^l(\bm r;B)-\langle M_z^l(B)\rangle$}, \mbox{$\langle M_z^l(B)\rangle$}} is the spatial mean over \Rev{layer} \Rev{\mbox{$l$},} and $S(0) = 1$ by construction.}

\begin{figure*}[t!]
    \centering
    \includegraphics[width=\textwidth]{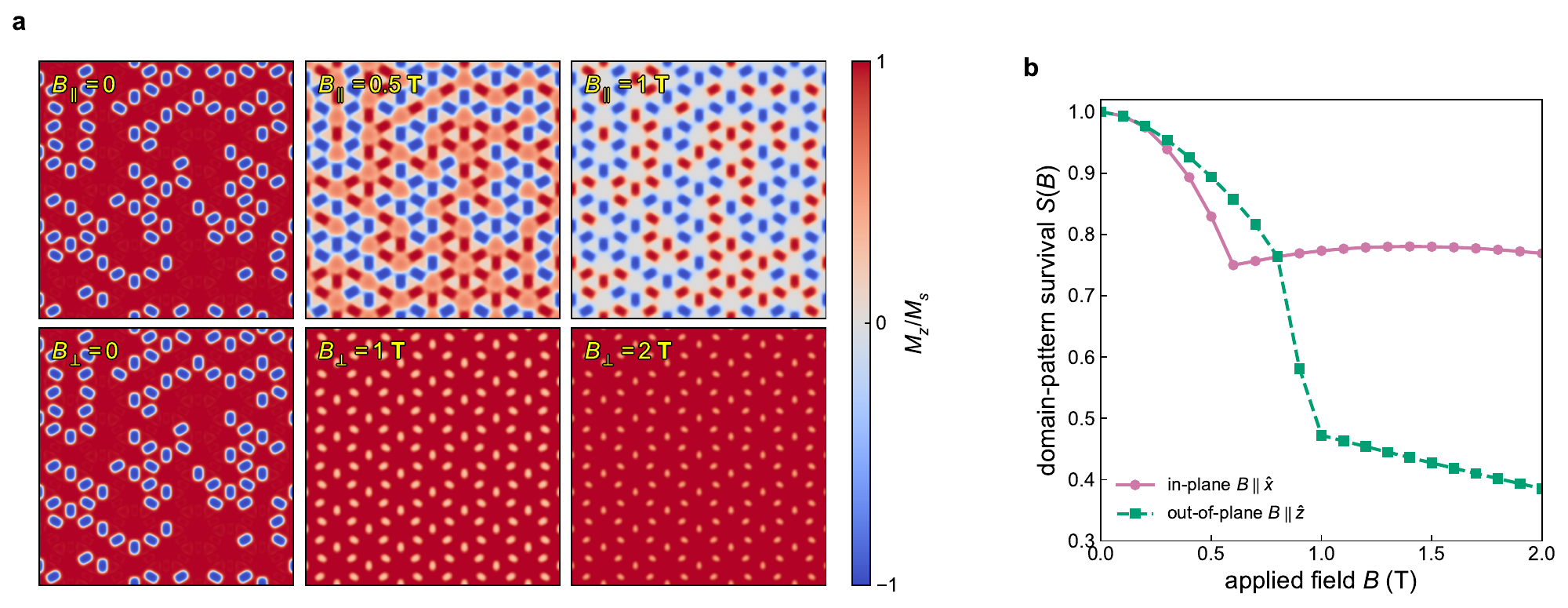}
    \caption{
    \textbf{Persistence of the domain pattern under an applied magnetic field.} \textbf{a}, Top-layer out-of-plane magnetization $M_z/M_s$ (red $\approx +1$, blue $\approx -1$) of a single relaxed configuration under an in-plane field $\bm{H} \parallel \hat{\bm{x}}$ (top row, at $B = 0$, $0.5$, and $1.0$~T) and an out-of-plane field $\bm{H} \parallel \hat{\bm{z}}$ (bottom row, at $B = 0$, $1.0$, and $2.0$~T). \textbf{b}, Domain-pattern survival $S(B)$ [Eq.~\eqref{eq:survival}], normalised to unity at zero field, versus applied field for the two orientations. The out-of-plane field (green) erodes the pattern (\Rev{\mbox{$S \approx 0.39$}} at $2$~T), whereas the in-plane field (pink) drives a partial collapse near \Rev{\mbox{$B \approx 0.6$}}~T to a field-resilient plateau \Rev{\mbox{$S \approx 0.77$}}.
    }
    \label{fig:field}
\end{figure*}

\R{Figure~\ref{fig:field} shows the outcome. Under an out-of-plane field (Fig.~\ref{fig:field}a, bottom row; green curve in Fig.~\ref{fig:field}b) the reversed domains are suppressed \Rev{strongly}, and $S$ decreases \Rev{sharply} \Rev{near} \Rev{\mbox{$0.9$}--\mbox{$1.0$}~T}, retaining \Rev{\mbox{$S \approx 0.39$}} even at $2$~T. An in-plane field (Fig.~\ref{fig:field}a, top row; pink curve) behaves qualitatively differently. $S$ \Rev{decreases to about \mbox{$0.75$}} near \Rev{\mbox{$B \approx 0.6$}}~T as the ferromagnetic background cants into the plane, before saturating at a robust plateau \Rev{\mbox{$S \approx 0.77$}} that is essentially field-independent up to $2$~T. Here $S$ is a normalised spatial-correlation coefficient, not a count of surviving patches; a large residual correlation indicates that much of the zero-field domain pattern persists even after the background has been canted away, but the patch-by-patch survival fraction is a distinct quantity that we do not equate with $S$.}

\section{Discussion}

Moir\'{e} magnetic chaos differs qualitatively from previously studied chaotic magnetic systems. Driven chaos in vortex oscillators, spin-torque nano-oscillators, and isolated skyrmions operates with only one to three macroscopic dynamical variables and requires sustained external forcing~\cite{Petit-Watelot2012, PhysRevLett.123.147701, PhysRevB.108.174441, PhysRevB.109.014422, PhysRevB.109.174420}, whereas moir\'{e} magnetic chaos is \Rev{transient} and high-dimensional, with its number of dynamical degrees of freedom set by the moir\'{e} supercell area. Several extensions are possible. Incorporating the interfacial Dzyaloshinskii--Moriya interaction, which arises from inversion-symmetry breaking at a substrate, converts the reversed domains into a lattice of magnetic skyrmions~\cite{Akram2021, Zheng2022, Kim2023, PhysRevB.108.174440, https://doi.org/10.1002/admi.202300188}, as observed recently in experiment~\cite{Wong2026, Kim2026}, opening the collective dynamics of coupled skyrmions beyond the chaos of isolated textures~\cite{PhysRevB.108.174441, PhysRevB.109.014422, PhysRevB.109.174420}. The phenomenon should likewise extend to other twisted van der Waals magnets~\cite{doi:10.1021/acs.nanolett.4c04582, Kim2025, Chen2026, doi:10.1073/pnas.2413326121}, most notably twisted CrSBr, extending chaotic moir\'{e} dynamics to twist-based spintronics more broadly~\cite{Yang2024, Chen2024, Chen2026}.

These predictions are directly accessible to experiment. Moir\'{e} magnetic chaos can be realized in twisted bilayer CrI\textsubscript{3}~\cite{doi:10.1126/science.abj7478, Xu2022} or its multilayer generalizations~\cite{Xie2022, Xie2023, Cheng2023, Li2024, Yang2024, Wong2026}, and detected by exploiting the inverse scaling of the domain size and superlattice period with twist angle: at $0.5^{\circ}$, for instance, these reach approximately $26$ and $80$~nm~\cite{Kim2023}, well within the reach of scanning quantum magnetometry~\cite{doi:10.1126/science.abj7478, Wong2026}, which can image the chaotically selected domain pattern directly. Because the pattern is \Rev{visible} \Rev{at} \Rev{low} temperature and \Rev{under} applied field (Figs.~\ref{fig:thermal_locking} and~\ref{fig:field}), a single cryogenic measurement \Rev{can} read out \Rev{a} \Rev{metastable} \Rev{configuration}.

In summary, we have demonstrated moir\'{e} magnetic chaos in twisted bilayer CrI\textsubscript{3}, an \Rev{transient}, high-dimensional sensitivity to initial conditions that emerges during relaxation without any external drive. Three lines of evidence \Rev{support} \Rev{this} \Rev{interpretation}: \Rev{positive} \Rev{finite-time} Lyapunov \Rev{exponents} \Rev{during} \Rev{relaxation} \Rev{and} \Rev{amplitude-dependent} \Rev{final-state} \Rev{sensitivity}; a per-patch Shannon entropy at \Rev{\mbox{$99.99\%$}} of its maximum across \Rev{\mbox{$10000$}} quenches; and pairwise domain correlations statistically indistinguishable from independence.

The extensive, \Rev{transient} character of moir\'{e} magnetic chaos suggests concrete advantages for unconventional computing. A single thermal quench \Rev{selects} a \Rev{metastable} \Rev{configuration}. In contrast to conventional chaotic devices, in which disorder is a parasitic source of irreproducibility, here the periodic moir\'{e} order is constitutive, so the chaos is at once reproducible and diagnostic of the twist engineering itself. These features make twisted van der Waals magnets suitable for hardware random-number generation and related unconventional-computing applications.

\section{Methods}

\subsection{Micromagnetic simulations.}

The magnetization dynamics is simulated by solving the LLG equation:
\begin{equation}
    \frac{d\bm{m}_i}{dt} = -\gamma \bm{m}_i \times \bm{H}_{i}^\textrm{eff} + \alpha \bm{m}_i \times \frac{d\bm{m}_i}{dt},
\end{equation}
where $\bm{m}_i$ is the unit-magnitude magnetization at simulation cell $i$, $\bm{H}_i^{\mathrm{eff}} = \partial E_{\mathrm{tot}} / \partial \bm{m}_i$ is the effective field obtained from the total energy $E_{\mathrm{tot}}$, $\gamma \approx 1.76 \times 10^{11}\,\mathrm{s}^{-1}\,\mathrm{T}^{-1}$ is the gyromagnetic ratio, and $\alpha = 0.02$ is the dimensionless Gilbert damping constant. Simulations use the mumax\textsuperscript{+} solver~\cite{Moreels2026}. The simulation grid consists of $588 \times 595 \times 1$ cells ($4 \times 7$ moir\'{e} unit cells, each $147 \times 85$ sites), with cell dimensions $\Delta x = \Delta y = 0.4525$~nm and $\Delta z = 0.67$~nm, an interlayer separation of $a = 0.67$~nm, and a total lateral size of $266 \times 269$~nm$^2$. Periodic boundary conditions ($10 \times 10$ repetitions) are applied in-plane. \Rev{For Figs.~\ref{fig:lyapunov}--\ref{fig:field}, the} sublattice parameters are the continuum counterparts of the spin-Hamiltonian constants of Sec.~\ref{sec:model}, namely \Rev{\mbox{$M_s = 1.8053 \times 10^{5}$}}~A/m, \Rev{\mbox{$A_\text{ex} = 2.5348 \times 10^{-13}$}}~J/m, and \Rev{\mbox{$K_{u1} = 1.0159 \times 10^{5}$}}~J/m$^3$ ($\hat{\bm{z}}$ easy axis)\Rev{. Figure~1 retains the original values \mbox{$M_s=1.82\times10^5$}~A/m, \mbox{$A_{\mathrm{ex}}=3.12\times10^{-13}$}~J/m and \mbox{$K_{u1}=1.18\times10^5$}~J/m\mbox{$^3$}}. The interlayer coupling map $J_{\mathrm{inter}}(\bm{r})$, spanning $-0.47$ to $+0.50$~meV, is obtained from the first-principles spin Hamiltonian of Ref.~\cite{Kim2023SpinHam} (Fig.~\ref{fig:model}a), and its layer-resolved implementation is detailed in Sec.~\ref{sec:methods_numerical}. \Rev{Integration} is \Rev{adaptive,} \Rev{in} \Rev{single} \Rev{precision at} $T = 0$~K \Rev{unless otherwise stated}. The simulations reproduce ground-state domains and skyrmions consistent with previous studies\cite{Kim2023, Kim2024, PhysRevB.108.L100401, Lee_2024, Kim2025, Cadez_2026}.

\subsection{Damage-spreading protocol.}

A reference state is initialised with $m_z = 0.99$ and randomised in-plane components (\Rev{seed} \Rev{\mbox{$20260824$}}), and relaxed for 2~ns at $T = 0$~K (\Rev{adaptive tolerance \mbox{$10^{-6}$}}). Perturbed trajectories use an identical initial state \Rev{with} \Rev{a} \Rev{projected} \Rev{Gaussian} \Rev{tangent} \Rev{field (seed \mbox{$60500$}) scaled} to \Rev{per-spin RMS amplitude \mbox{$\varepsilon$}, added and normalized per spin}, with \Rev{10} values of $\varepsilon$ logarithmically spaced from \Rev{\mbox{$10^{-5}$}} to \Rev{\mbox{$10^{-2}$}}. After 2~ns relaxation, the RMS difference $\delta m$ and Hamming distance of \Rev{domain} bitstreams are computed. \Rev{The} \Rev{\mbox{$157$} interior AFM patches} are identified \Rev{and} assigned \Rev{bits} \Rev{as} \Rev{in} \Rev{the} \Rev{statistical} \Rev{analysis} \Rev{below}. The \Rev{amplification} \Rev{rate} follows from Eq.~\eqref{eq:lyapunov}\Rev{. Figure~\ref{fig:lyapunov}c uses the same initialization seed but tangent-direction seed \mbox{$1000$}, double precision, fixed \mbox{$6.25$}~fs steps and \mbox{$1$}~ps Benettin renormalization, with finite-time Lyapunov exponents averaged over \mbox{$20$}~ps}.

\subsection{Statistical analysis of domain configurations.}

We perform \Rev{10000} independent runs, each relaxed for 2~ns at $T = 0$~K. The initial states are not chosen arbitrarily. Every run starts from the same out-of-plane-polarised, domain-free ferromagnetic background and differs only in a random in-plane texture. Concretely, the in-plane components are drawn independently from a normal distribution, $m_x^{(0)}, m_y^{(0)} \sim \mathcal{N}(0,1)$, the out-of-plane component is fixed at $m_z^{(0)} = m_z^{\text{fixed}} = 0.99$, and the vector is normalised to unit length, $\bm{m}^{(0)} = (m_x^{(0)}, m_y^{(0)}, m_z^{(0)})/\lvert (m_x^{(0)}, m_y^{(0)}, m_z^{(0)}) \rvert$. Distinct random seeds therefore correspond to distinct random in-plane fluctuations of one and the same polarised initial state, so the ensemble isolates how the deterministic relaxation amplifies the in-plane fluctuations into different domain patterns. The $N = 157$ AFM patches are the interior AFM-coupled sites, identified as connected components of the $J_{\mathrm{inter}} > 0$ map ($\geq 50$ cells, boundary-touching sites excluded), and each patch is assigned a binary state from the sign of its mean $m_z$. Per patch we compute the reversal probability $p_{-1}^{(n)}$ and binary Shannon entropy $H_n$ (Eq.~\eqref{eq:shannon}). Inter-patch independence is assessed from the pairwise Hamming-distance distribution (compared with $\mathrm{Bin}(N,1/2)$) and from the Pearson correlation coefficients $\rho_{ij}$ over all patch pairs, tested against the $\mathcal{N}(0, 1/n_\text{runs})$ null with Bonferroni correction.

\R{\subsection{Heat sweep and domain-order analysis.}} \label{sec:methods_heatsweep}

\R{To probe the thermal robustness of the chaotically selected pattern (Fig.~\ref{fig:thermal_locking}), we \Rev{evolve} \Rev{the} \Rev{same} relaxed configuration \Rev{separately} \Rev{at each temperature} using the stochastic LLG solver with a Langevin field whose two-point correlator follows the fluctuation--dissipation theorem at the prescribed $T$. At each temperature \Rev{\mbox{$T = 0$}}--$29$~K (in $1$~K steps) the same \Rev{relaxed} configuration \Rev{(ensemble run \mbox{$1$})} is evolved \Rev{for} \Rev{\mbox{$2$}~ns with adaptive integration and \mbox{$10$}}~ps \Rev{output cadence}, and the top-layer out-of-plane magnetization is sampled over the \Rev{observation} window $t \in [1,2]$~ns. The domain order $\mathcal{O}(T)$ [Eq.~\eqref{eq:order}] is computed as the spatial standard deviation of the time-averaged map $\overline{m}_z(\bm{r})$ [Eq.~\eqref{eq:timeavg}], with the $T = 0$ reference \Rev{evaluated} \Rev{over} the \Rev{same} \Rev{\mbox{$101$}-frame} \Rev{window}.}

\R{\subsection{Layer-resolved implementation and exchange-map calibration.}} \label{sec:methods_numerical}

\R{The bilayer is mapped onto a single two-dimensional grid whose two co-located magnetic sublattices represent the top and bottom CrI\textsubscript{3} layers, separated by $a = 0.67$~nm; the intralayer constants above are shared by both sublattices, so each layer behaves as a ferromagnetic monolayer. The stacking-dependent interlayer coupling of Eq.~\eqref{eq:spinH} is entered as a cell-resolved on-site exchange between the two co-located moments, whose local sign encodes ferromagnetic ($J_{\mathrm{inter}} < 0$) or antiferromagnetic ($J_{\mathrm{inter}} > 0$) stacking. Any spatially-homogeneous, neighbour-to-neighbour interlayer term is set to zero, so the interlayer interaction is carried purely by this vertical on-site coupling. The single moir\'{e}-unit-cell map is tiled $4 \times 7$ to span the full $588 \times 595$ grid.}

\R{The mumax\textsuperscript{+} solver~\cite{Moreels2026} integrates the deterministic LLG equation with a high-order adaptive Runge--Kutta--Fehlberg scheme and \Rev{includes} a \Rev{thermal} \Rev{field at finite temperature}. The cell volume \Rev{is \mbox{$V_\text{cell}=\Delta x\,\Delta y\,\Delta z=1.37\times10^{-28}$}}~m$^3$. The atomistic-to-continuum mapping of the interlayer exchange uses a calibration factor \Rev{\mbox{$\zeta = -1.3104 \times 10^{-13}$}}~J\,m$^{-1}$ per meV, converting the unit-cell map $J_{\mathrm{inter},[\text{meV}]}(\bm{r})$ of Ref.~\cite{Kim2023SpinHam} to the local micromagnetic coupling $\zeta \cdot J_{\mathrm{inter},[\text{meV}]}(\bm{r})$. \Rev{Figure~1} \Rev{retains} the \Rev{original} factor \Rev{\mbox{$\zeta=-5.38\times10^{-13}$}}~J\,m$^{-1}$ per meV.}

\section*{Acknowledgements}

K. K. was supported by an appointment to the JRG Program at the APCTP through the Science and Technology Promotion Fund and Lottery Fund of the Korean Government, the Korean Local Governments (Gyeongsangbuk-do Province and Pohang City), and the National Research Foundation of Korea (NRF) funded by the Korean government (Ministry of Science and ICT, MSIT) (No. RS-2026-25499525). This study was supported by the National Research Council of Science and Technology (NST) grant by the Korea government (MSIT) (Grant No. GTL24042-000) and the KIST Institutional Program (Grant No. 26E0031).

\section*{Data availability}

The data generated in this study are available from the corresponding author upon reasonable request.

\section*{Code availability}

The micromagnetic simulations were performed using the mumax\textsuperscript{+} solver, which is publicly available. Custom analysis scripts are available from the corresponding author upon reasonable request.

\section*{Author contributions}

G.P.\ performed the micromagnetic simulations and data analysis. O.L.\ contributed to the discussion. K.-M.K.\ conceived and supervised the project. All authors contributed to writing the manuscript.

\section*{Competing interests}

The authors declare no competing interests.

\input{main.bbl}
\clearpage\section*{Table of Contents}

\noindent\textbf{Emergence of moir\'{e} magnetic chaos in twisted bilayer CrI\textsubscript{3}}

The study of magnetic chaos has traditionally focused on macroscopic variables under external driving. The domains are stabilized by a characteristic interlayer exchange frustration, which supplies the multiple dynamical degrees of freedom required for \Rev{transient chaos}. Statistical analysis further reveals that the resulting domain configurations are stochastic and pairwise uncorrelated.

\begin{center}
\begin{minipage}[c][50mm][c]{55mm}\centering
\includegraphics[width=55mm,height=50mm,keepaspectratio]{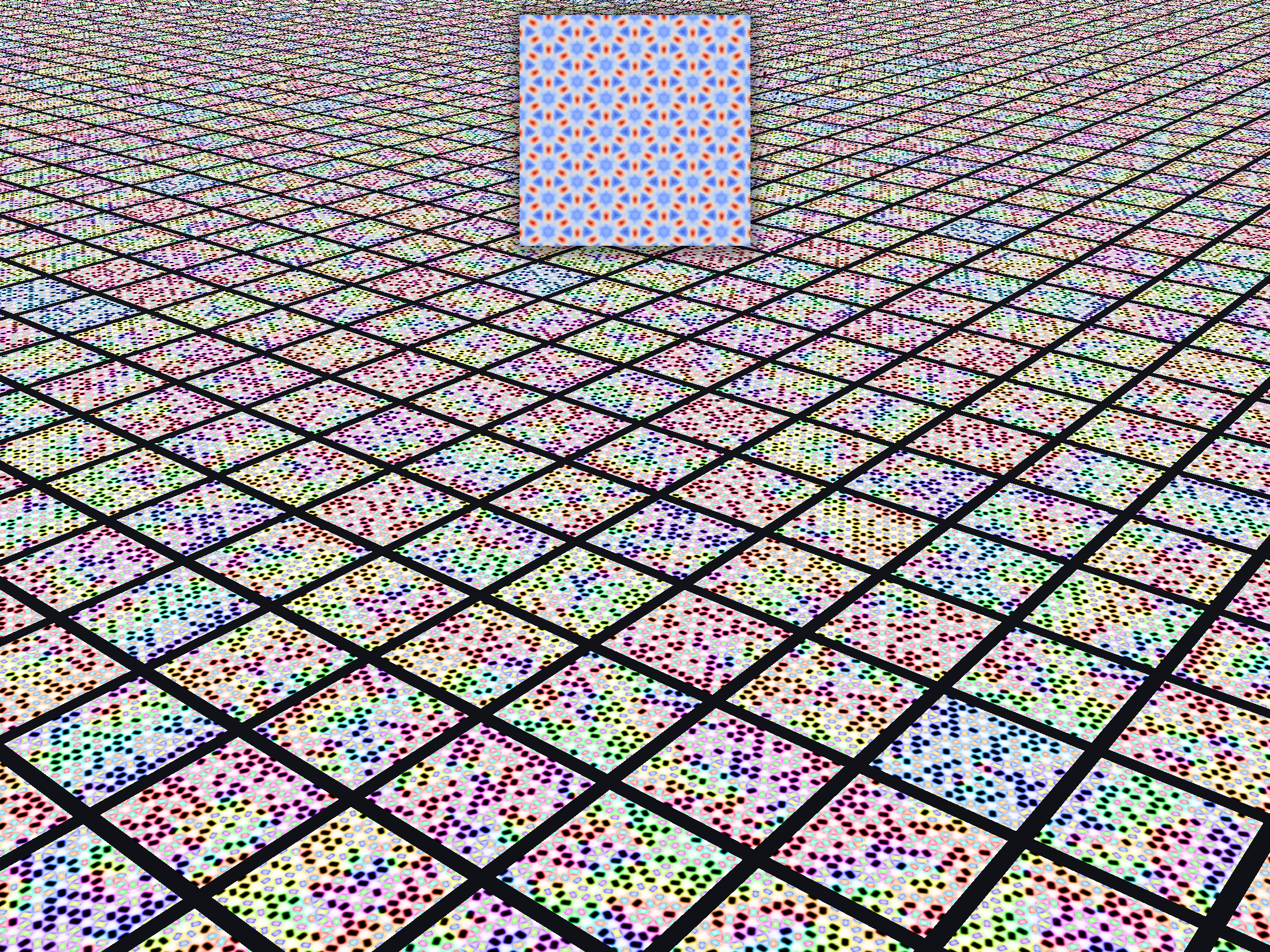}
\end{minipage}
\end{center}

\clearpage
\setcounter{section}{0}
\setcounter{subsection}{0}
\setcounter{figure}{0}
\setcounter{table}{0}
\setcounter{equation}{0}
\setcounter{footnote}{0}
\setcounter{page}{1}
\renewcommand{\thesection}{S\arabic{section}}
\renewcommand{\thefigure}{S\arabic{figure}}
\renewcommand{\theequation}{S\arabic{equation}}
\renewcommand{\thepage}{S\arabic{page}}
\renewcommand{\theHsection}{supp.\arabic{section}}
\renewcommand{\theHfigure}{supp.\arabic{figure}}
\renewcommand{\theHequation}{supp.\arabic{equation}}
\begin{center}
\textbf{\large Supplementary Information}\\[6pt]
\textbf{Emergence of moir\'{e} magnetic chaos in twisted bilayer CrI\textsubscript{3}}\\[2pt]
Gyuyoung Park, OukJae Lee, Kyoung-Min Kim
\end{center}
\vspace{6pt}
\section{Skyrmion versus bubble classification of the reversed domains.} \label{sec:skbubble}

The reversed domains that nucleate at the AFM patches fall into two topological classes, distinguished by the skyrmion (topological) charge $Q = \frac{1}{4\pi}\int \bm{m}\cdot(\partial_x\bm{m}\times\partial_y\bm{m})\,\mathrm{d}^2r$ integrated over each domain. Fig.~\ref{fig:skbubble} shows the in-plane magnetization of a representative example of each class in both layers. A skyrmion-like domain (top row) exhibits a continuous winding of the in-plane angle around its core and carries a near-integer charge $Q \approx +1$ in the layer hosting the reversal, while the complementary region in the opposite layer is topologically trivial ($Q \approx 0$). A bubble-like domain (bottom row) shows the same out-of-plane reversal but with $Q \approx 0$ in both layers, its in-plane texture lacking the net winding. Both types occur among the chaotically selected domains, and the skyrmionic character of the former is confirmed independently by the eigenmode spectroscopy of Sec.~\ref{sec:suppl_modes}.

\begin{figure*}[b!]
    \centering
    \includegraphics[width=0.6\textwidth]{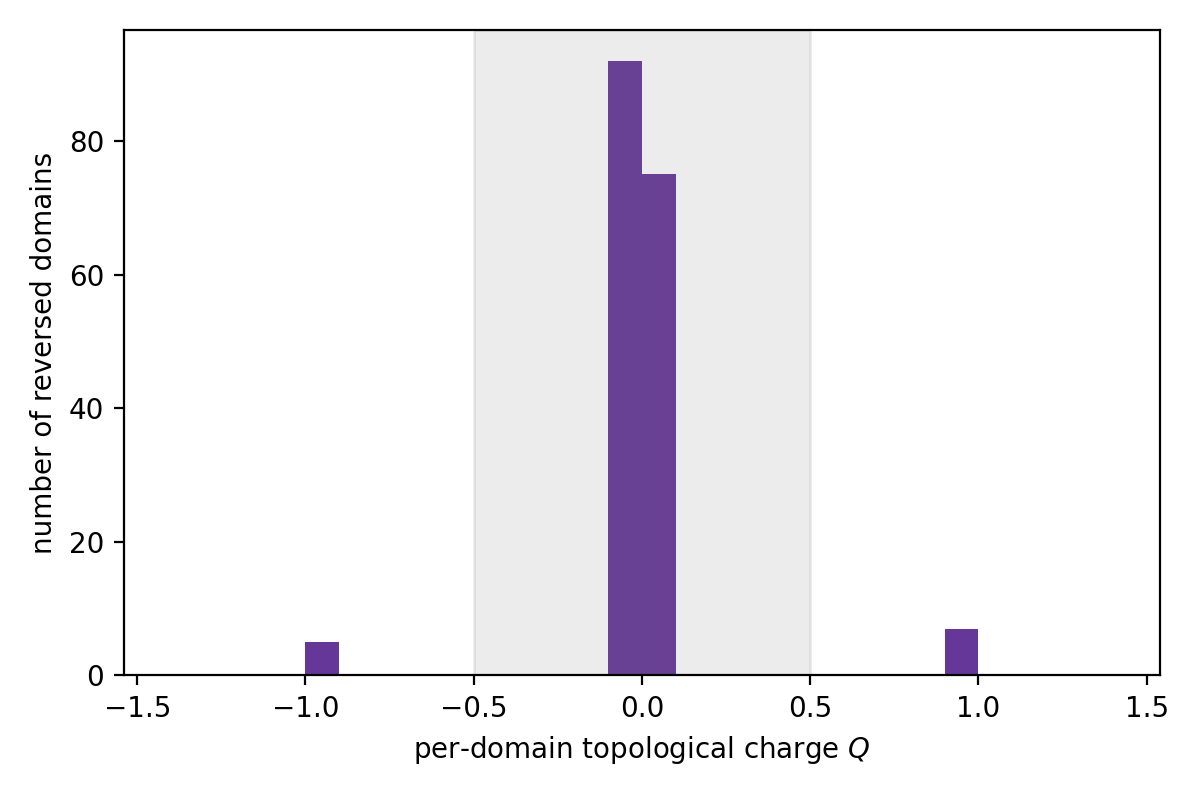}
    \caption{%
    \textbf{Topological-charge distribution of the reversed domains.} Histogram of the per-domain topological charge $Q$ over all $179$ reversed domains of the reference configuration (top- and bottom-layer domains combined), computed from the lattice solid-angle charge density on the dilated domain footprint. The distribution is bimodal, with skyrmion-like domains at $|Q| \approx 1$ and bubble-like domains at $Q \approx 0$ (grey band, $|Q| < 0.5$). Applying $|Q| = 0.5$ gives ${\approx}7\%$ skyrmion-like and ${\approx}93\%$ bubble-like.%
    }
    \label{fig:qdist}
\end{figure*}

\begin{figure*}[t!]
    \centering
    \includegraphics[width=0.72\textwidth]{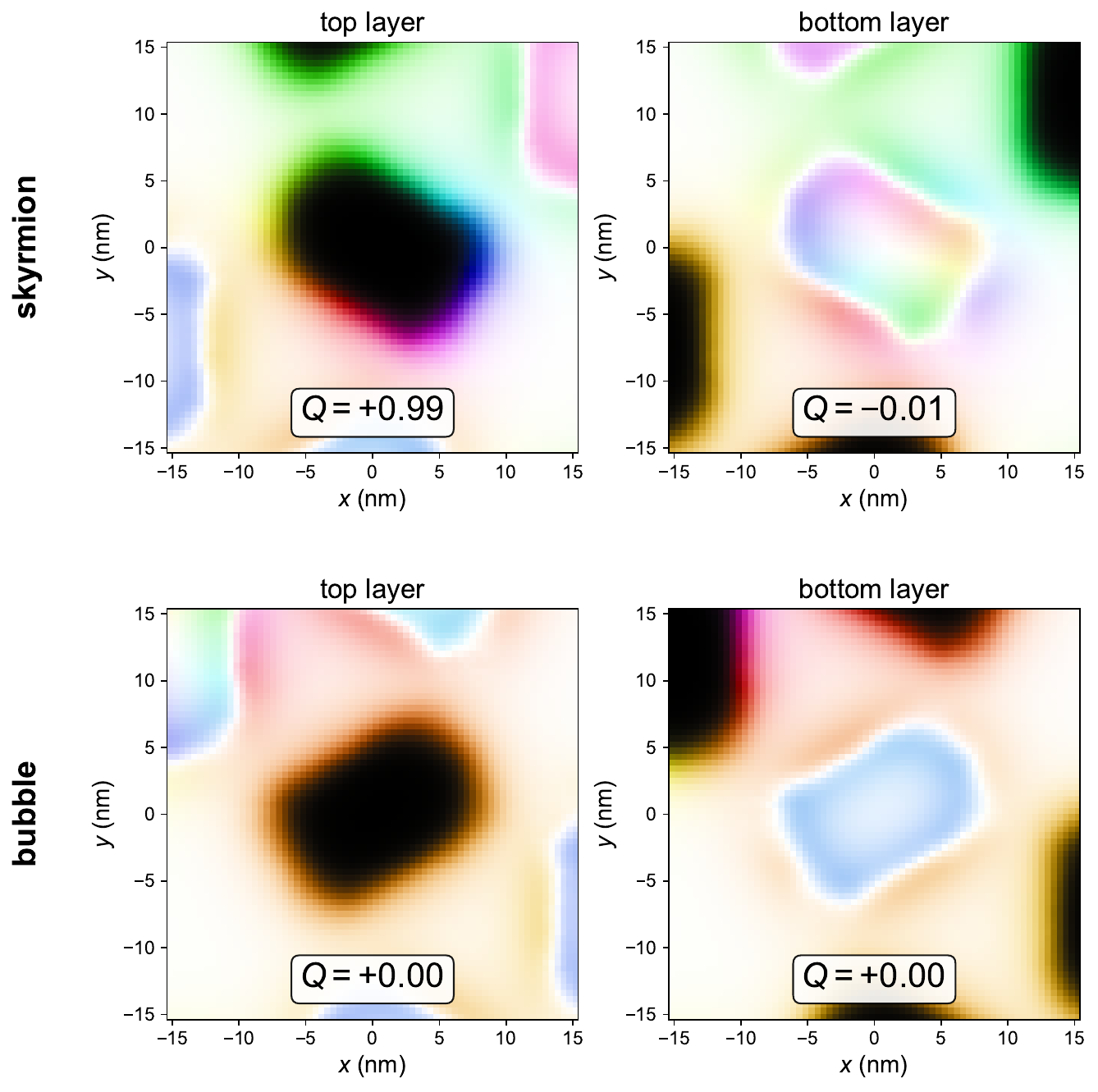}
    \caption{%
    \textbf{Skyrmion versus bubble classification of the reversed domains.} In-plane magnetization (hue encodes the in-plane angle $\mathrm{atan2}(m_y, m_x)$, brightness the out-of-plane component) of a representative skyrmion-like domain (top row) and bubble-like domain (bottom row), each shown in both the top and bottom CrI\textsubscript{3} layers over a ${\sim}31 \times 31$~nm window centred on the domain. Each panel is annotated with the topological charge $Q$ integrated over the domain: the skyrmion-like domain carries $Q \approx +1$ in the reversed layer and $Q \approx 0$ in the other, whereas the bubble-like domain has $Q \approx 0$ in both.%
    }
    \label{fig:skbubble}
\end{figure*}

To quantify how the two classes are distributed, we compute the topological charge $Q$ of every reversed domain in the \Rev{original} configuration \Rev{of main-text Fig.~\ref{fig:model}b} from the lattice solid-angle charge density integrated over the (dilated) domain footprint. The distribution (Fig.~\ref{fig:qdist}) is sharply bimodal: $|Q|$ reaches ${\approx}1$ for the skyrmion-like domains and clusters near $0$ for the bubble-like domains, with essentially no domains in between. With a threshold $|Q| = 0.5$, $12$ of the $179$ reversed domains (${\approx}7\%$, counting both layers) are skyrmion-like and the remaining ${\approx}93\%$ are bubble-like. Bubble-like domains therefore dominate the reference pattern, and the skyrmion statements of this section and of Sec.~\ref{sec:suppl_modes} refer specifically to this ${\approx}7\%$ skyrmion-like subset.

\clearpage

\section{Energy density of states and the origin of the high metastability.} \label{sec:dos}

\begin{figure*}[t!]
    \centering
    \includegraphics[width=0.62\textwidth]{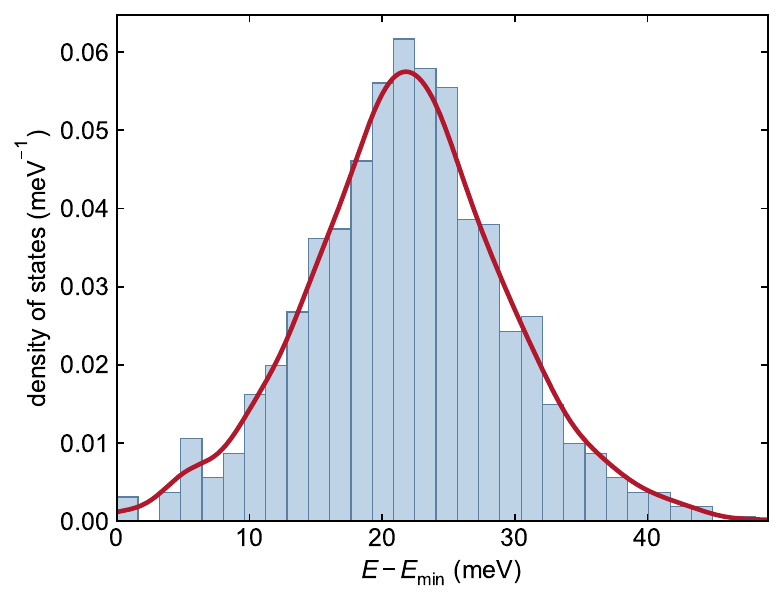}
    \caption{%
    \textbf{Energy density of states of the metastable ensemble.} Distribution of the re-evaluated $0$~K total energies of the $1000$-seed unbiased quench ensemble, plotted relative to the minimum energy $E_{\min}$. The whole manifold spans only $E_{\max} - E_{\min} \approx 48$~meV ($\Delta E / |E| \approx 1.8 \times 10^{-4}$). The red curve is a smooth kernel-density estimate of the density of states. The extremely narrow, near-Gaussian density of states (quantified in the text) reflects a sum of many nearly independent per-patch energies and underlies the high metastability.%
    }
    \label{fig:dos}
\end{figure*}

To quantify the near-degeneracy of the metastable manifold, we re-evaluate the $0$~K total energy of every configuration in the \Rev{original} $1000$-seed ensemble with the same Hamiltonian as the relaxation runs (static evaluation, no dynamics). Fig.~\ref{fig:dos} shows the resulting density of states. The domain-carrying configurations occupy an extremely narrow energy window, $E_{\max} - E_{\min} \approx 48$~meV for the whole $588 \times 595$ bilayer, corresponding to a relative spread $\Delta E / |E| \approx 1.8 \times 10^{-4}$, i.e.\ a mean energy spacing of ${\approx}0.31$~meV per patch when the full ensemble range is divided by $N = 157$. We emphasise that this is a spacing across the sampled ensemble, not a directly computed single-patch flip energy or barrier. A single anomalous run that relaxed to a nearly domain-free ferromagnetic state lies far above this window and is excluded from the fit.

The density of states is very well described by a Gaussian (standard deviation $\sigma \approx 7.5$~meV; Kolmogorov--Smirnov statistic $0.03$, skewness $0.05$, excess kurtosis $0.26$). The Gaussian shape is consistent with a central-limit-theorem picture in which the total energy is a sum of many nearly independent per-patch contributions. Because the ensemble energy spread, ${\sim}0.3$~meV per patch, is far smaller than the intralayer exchange and anisotropy that stabilise each reversed domain, no configuration within the sampled $1000$-state ensemble is energetically singled out. This flat, near-degenerate landscape underlies the high metastability that enables the chaotic, unbiased selection of domain patterns.

\section{Enumeration of the AFM patches.} \label{sec:patch_index}

\begin{figure*}[t!]
    \centering
    \includegraphics[width=\textwidth]{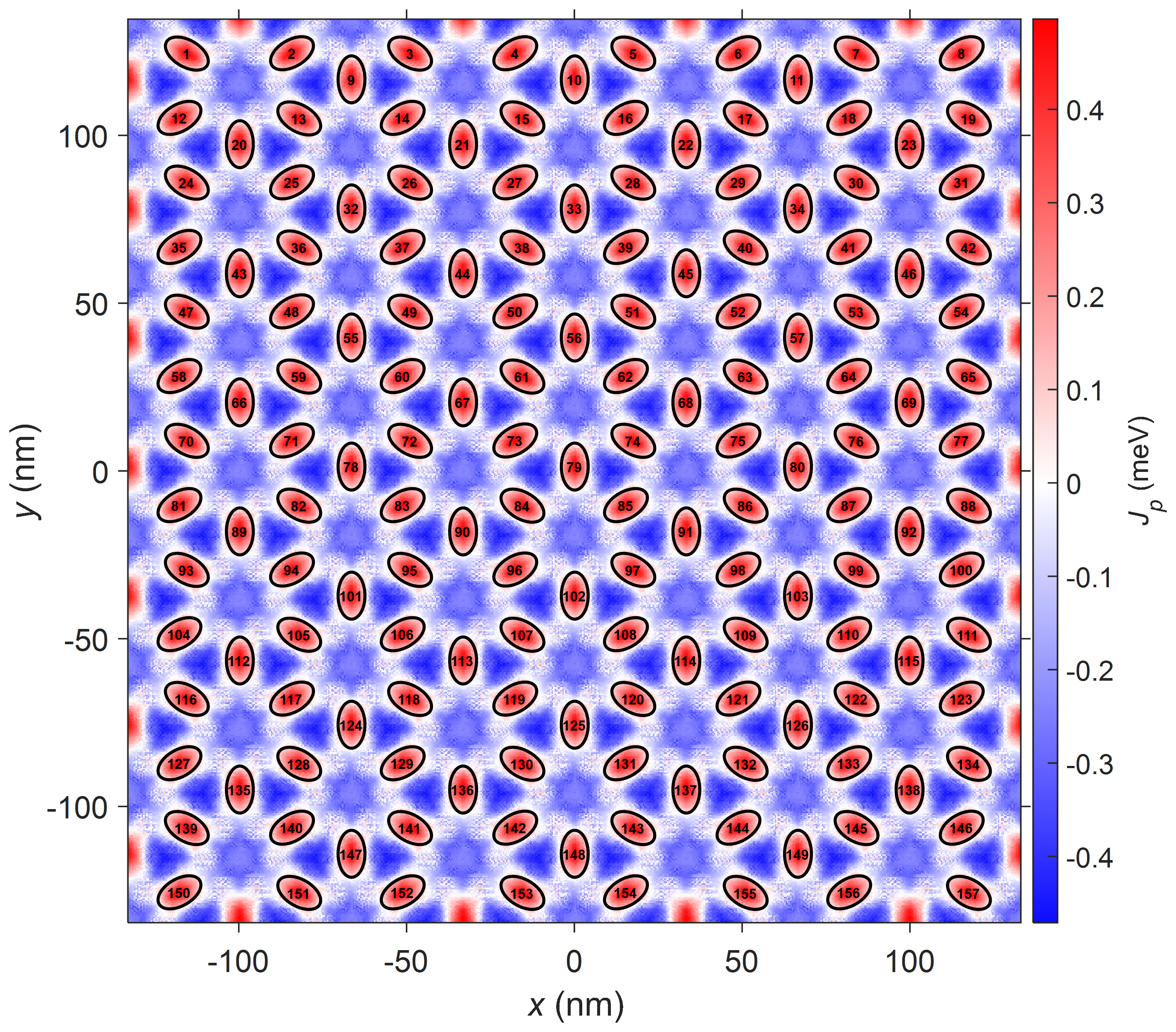}
    \caption{%
    \textbf{Enumeration of the $N = 157$ AFM patches.} The interlayer-exchange map $J_{\mathrm{inter}}(\bm{r})$ (colour scale, in meV; the same map as Fig.~\ref{fig:model}a of the main text) with every interior AFM patch ($J_{\mathrm{inter}} > 0$) outlined and assigned a fixed index $n = 1, \ldots, 157$. Patches touching the simulation boundary are excluded from the enumeration. This indexing is used consistently for the binary-string, entropy, and pairwise-correlation analyses of the main text.%
    }
    \label{fig:patch_index}
\end{figure*}

The statistical analysis of the main text records each relaxed configuration as a binary string over a fixed set of AFM patches. The patches are defined directly from the interlayer-exchange map: the AFM-coupled sites are the connected components of the region $J_{\mathrm{inter}}(\bm{r}) > 0$, keeping only components of at least $50$ cells and discarding any that touch the simulation boundary. This yields $N = 157$ interior patches. Each patch is assigned a fixed index $n = 1, \ldots, 157$, ordered by the position of its centroid from top to bottom and, within each row, from left to right, as shown in Fig.~\ref{fig:patch_index}. The same indexing is used throughout for the per-patch reversal probability, the Shannon entropy, and the pairwise correlation analysis, so that bit $n$ refers to the same physical patch in every run.

\section{Shannon and correlation entropy of the domain ensemble.} \label{sec:entropy}

\begin{figure*}[t!]
    \centering
    \includegraphics[width=0.86\textwidth]{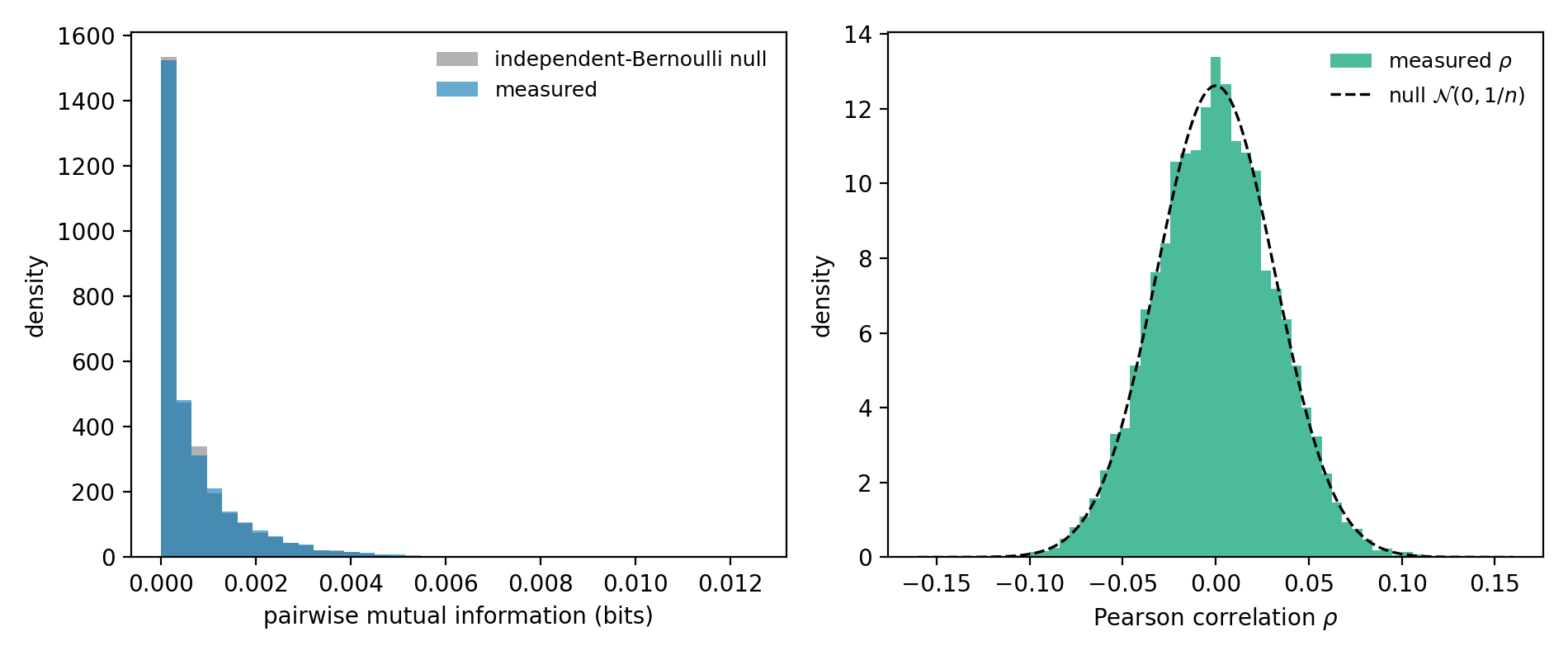}
    \caption{%
    \textbf{Pairwise mutual information and Pearson correlation of the $157$ occupancy bits.} \textbf{Left}, distribution of the pairwise mutual information $I(X_i;X_j)$ over all $12{,}246$ patch pairs (blue) versus an independent-Bernoulli null obtained by permuting each bit column independently (grey); the two coincide, so no dependence is detected beyond the finite-sample floor. \textbf{Right}, distribution of the Pearson coefficient $\rho$ (green) versus the finite-sample independence null $\mathcal{N}(0, 1/n)$ (dashed), $n = 1000$.%
    }
    \label{fig:mi}
\end{figure*}

\begin{figure*}[t!]
    \centering
    \includegraphics[width=0.92\textwidth]{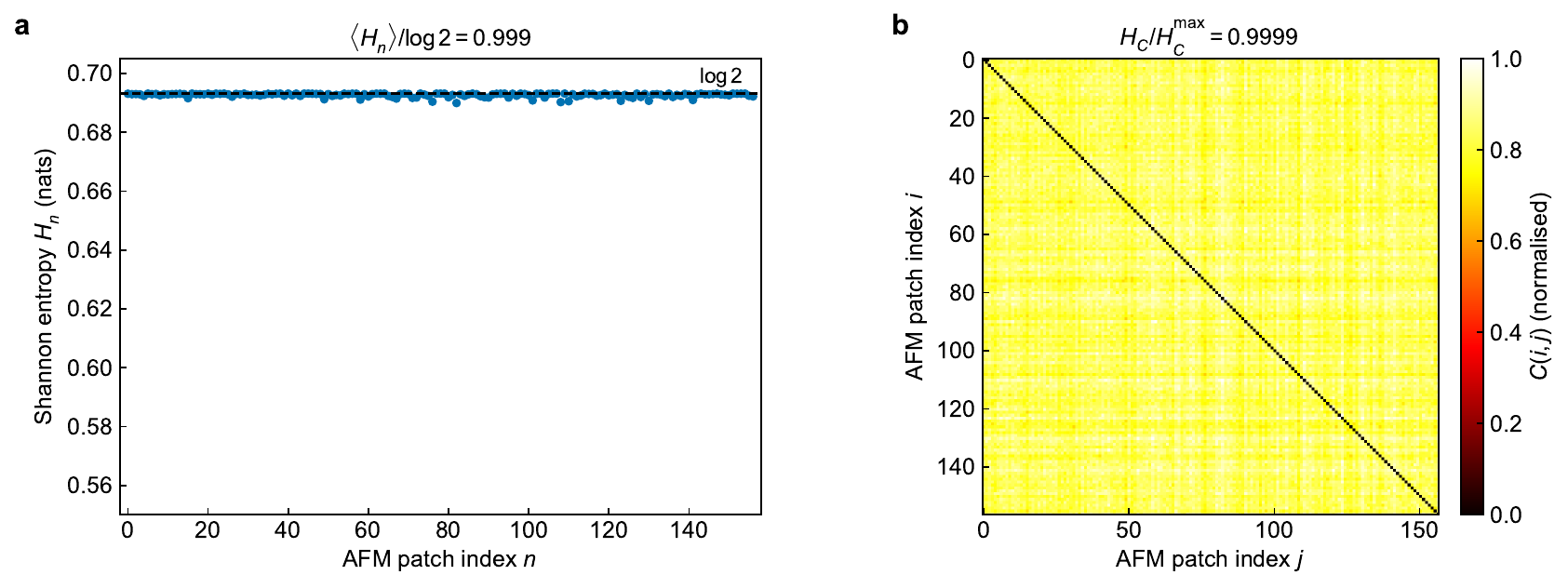}
    \caption{%
    \textbf{Shannon and correlation entropy of the $1000$-seed ensemble.} \textbf{a}, Per-patch binary Shannon entropy $H_n$ (nats) for all $N = 157$ AFM patches; every value lies just below the maximum $\log 2$ (dashed line), giving $\langle H_n\rangle/\log 2 = 0.999$. \textbf{b}, Normalised pairwise co-occurrence matrix $C(i,j)$ (diagonal excluded); its near-uniform structure corresponds to a correlation entropy $H_C/H_C^{\max} = 0.9999$, indicating no heterogeneity in the pairwise co-occurrence rates (this measure alone does not establish statistical independence; see text and the Pearson-coefficient test of the main text).%
    }
    \label{fig:entropy}
\end{figure*}

\Rev{For} \Rev{comparison} \Rev{with} the \Rev{main-text} \Rev{ensemble,} \Rev{we summarise} the $1000$-seed ensemble through two entropies, reported per patch and per pair in Fig.~\ref{fig:entropy}. The per-patch binary Shannon entropy, $H_n = -p_{-1}^{(n)}\log p_{-1}^{(n)} - (1-p_{-1}^{(n)})\log(1-p_{-1}^{(n)})$, lies within $0.1\%$ of its maximum $\log 2$ for every patch ($\langle H_n\rangle/\log 2 = 0.999$), so each patch behaves as an unbiased binary variable. The normalised pairwise co-occurrence matrix $C(i,j)$ is featureless away from the excluded diagonal, and its correlation entropy reaches $H_C/H_C^{\max} = 0.9999$ with $H_C^{\max} = \log[N(N-1)]$. We stress, however, that $H_C$ measures only the \emph{heterogeneity} of the pairwise co-occurrence rates and does not by itself establish statistical independence: an ensemble in which all patches reverse together ($X_1 = \cdots = X_N$) would also yield a uniform $C(i,j)$ and $H_C/H_C^{\max} \approx 1$ while being perfectly correlated. The evidence for the absence of pairwise correlation therefore rests on \Rev{a} direct Pearson-coefficient test of the \Rev{original} \Rev{\mbox{$1000$}-run ensemble} (mean $\rho \approx 0$, $|\rho| \lesssim 0.13$, no pair surviving Bonferroni correction), which we summarise here as: no statistically detectable pairwise correlations are present within the resolution of the $1000$-run ensemble. As a direct test of pairwise dependence that does not rely on $H_C$, we compute the mutual information $I(X_i;X_j)$ of every patch pair over the $1000$-run ensemble and compare it with an independent-Bernoulli null obtained by permuting each bit column independently (Fig.~\ref{fig:mi}). The measured mutual-information distribution is indistinguishable from the null (means $7.4 \times 10^{-4}$ versus $7.2 \times 10^{-4}$~bits; only $1.1\%$ of pairs exceed the null $99$th percentile, as expected for independent variables), and the Pearson coefficients ($\langle\rho\rangle = -0.0003$, $|\rho| \leq 0.13$) fall on the finite-sample independence null $\mathcal{N}(0, 1/n)$. No statistically detectable pairwise dependence is present within the resolution of the $1000$-run ensemble; mutual independence, however, is not thereby established.

\section{Skyrmion eigenmodes of the moir\'{e} domains.} \label{sec:suppl_modes}

A subset of the reversed domains stabilized at the AFM patches (Fig.~\ref{fig:model}) are skyrmion-like textures ($Q \approx +1$, Sec.~\ref{sec:skbubble}). To probe the internal dynamics of these skyrmion-like domains we perform picosecond-cadence micromagnetic spectroscopy at $T = 2$~K, in which the modes are excited by the thermal fluctuations of the stochastic (Langevin) LLG dynamics at that temperature. Starting from a relaxed multi-domain configuration, we record the local magnetization every $2$~ps for $20$~ns (sampling Nyquist frequency $250$~GHz, spectral resolution $\Delta f = 50$~MHz) and Fourier-transform the time series, decomposing the rim magnetization of each domain into azimuthal channels indexed by $m = 0, 1, 2, \ldots$ around its core.

\begin{figure*}[t!]
    \centering
    \includegraphics[width=0.9\textwidth]{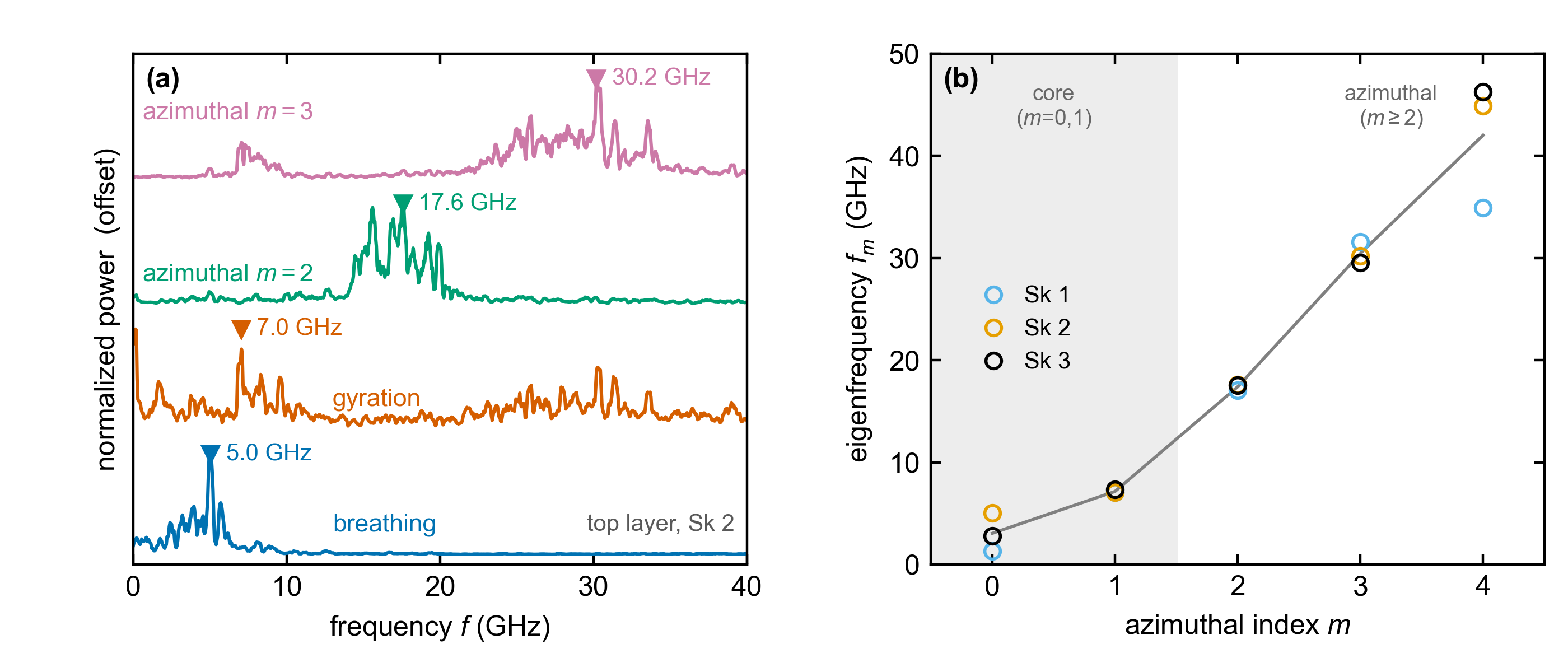}
    \caption{%
    \textbf{Skyrmion eigenmodes of the moir\'{e} domains ($T = 2$~K).} \textbf{a}, Power spectra of the rim out-of-plane magnetization of a representative domain, decomposed into azimuthal channels, revealing the breathing ($m = 0$), gyrotropic ($m = 1$), and higher azimuthal ($m = 2$--$4$) modes; the dominant peak of each channel is marked. \textbf{b}, Eigenfrequency $f_m$ versus azimuthal index $m$ for three independent domains; the near-collapse of the three datasets identifies the modes as intrinsic to the moir\'{e}-pinned skyrmions. Spectra obtained from a $20$~ns, $2$~ps-cadence run (Nyquist $250$~GHz, $\Delta f = 50$~MHz).%
    }
    \label{fig:suppl_modes}
\end{figure*}

\begin{figure*}[t!]
    \centering
    \includegraphics[width=\textwidth]{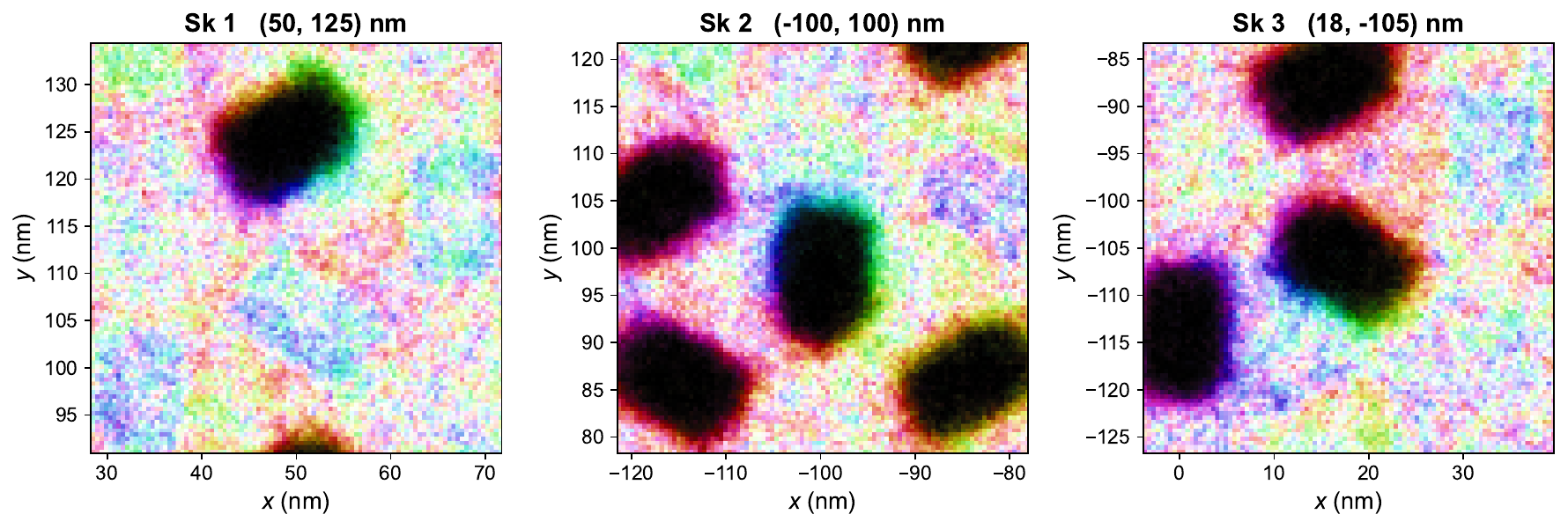}
    \caption{%
    \textbf{Locations of the three analysed skyrmions (Sk 1--3, $T = 2$~K, $t = 0$).} In-plane magnetization (HSV encoding, as in Fig.~\ref{fig:skbubble}) in a ${\sim}44 \times 44$~nm window centred on each of the three top-layer skyrmions whose eigenmodes are reported in Fig.~\ref{fig:suppl_modes}b, with the centre coordinate given above each panel.%
    }
    \label{fig:skloc}
\end{figure*}

As shown in Fig.~\ref{fig:suppl_modes}, the analysed skyrmion-like domains support the hierarchy of eigenmodes characteristic of magnetic skyrmions, namely a breathing mode ($m = 0$, ${\approx}3$--$5$~GHz), a gyrotropic (translational) mode ($m = 1$, ${\approx}7$~GHz), and higher-order azimuthal modes ($m \geq 2$; the $m = 2$ and $3$ peaks fall at ${\approx}17$ and ${\approx}30$~GHz, with a weaker $m = 4$ feature also resolved in Fig.~\ref{fig:suppl_modes}b). The eigenfrequencies measured for the three independent domains labelled Sk~1--3, whose locations are shown in Fig.~\ref{fig:skloc}, nearly collapse onto a single $f(m)$ ladder, identifying these as intrinsic skyrmion modes rather than configuration-specific features. This confirms that the analysed skyrmion-like subset consists of genuine moir\'{e}-pinned skyrmionic textures whose internal dynamics reproduce those of well-isolated magnetic skyrmions; the fraction of the full domain population that is skyrmionic is quantified separately (${\approx}7\%$; Sec.~\ref{sec:skbubble}).

\clearpage
\section{Null control: shuffled interlayer coupling.} \label{sec:null}

To test whether the chaotic, unbiased domain selection is specific to the moir\'{e}-structured interlayer frustration, we repeat the relaxation protocol ($100$ runs) with the interlayer-coupling map $J_{\mathrm{inter}}(\bm{r})$ spatially shuffled, which destroys the moir\'{e} spatial coherence while preserving the value distribution of the coupling. The outcome is unambiguous: with the shuffled map no reversed domains form at all---the reversed-area fraction is zero in every run ($0/100$ runs show any reversal exceeding $0.5\%$ of the area), compared with a mean reversed-area fraction of $0.14$ for the moir\'{e} map---and the AFM-coupled region no longer organizes into extended patches, leaving only $2$ interior clusters of $\geq 50$ cells versus $157$ for the moir\'{e} map. The metastable-domain manifold, and therefore the chaotic selection it enables, thus requires the spatially coherent moir\'{e} frustration and is not a generic consequence of coupling-strength disorder. A spatially uniform-coupling control is left for completeness.

\section{\Rev{Finite-time} Lyapunov \Rev{exponents} and the transient-chaos scenario.}\label{sec:lyapunov}

\Rev{The} \Rev{finite-time Lyapunov calculation of main-text Fig.~\ref{fig:lyapunov}c uses the Benettin finite-separation procedure of Eq.~\eqref{eq:local_growth}. Both trajectories use the same initialization seed \mbox{$20260824$}, and a projected Gaussian tangent direction uses seed \mbox{$1000$}. The state arrays, assignments and normalization are retained in double precision. Every \mbox{$\tau=1$}~ps we record \mbox{$d_k$}, rescale the separation to \mbox{$\delta_0$} and normalize each spin. The calculation uses one initial state and one initial tangent direction.}

\begin{figure*}[t!]
    \centering
    \includegraphics[width=\textwidth]{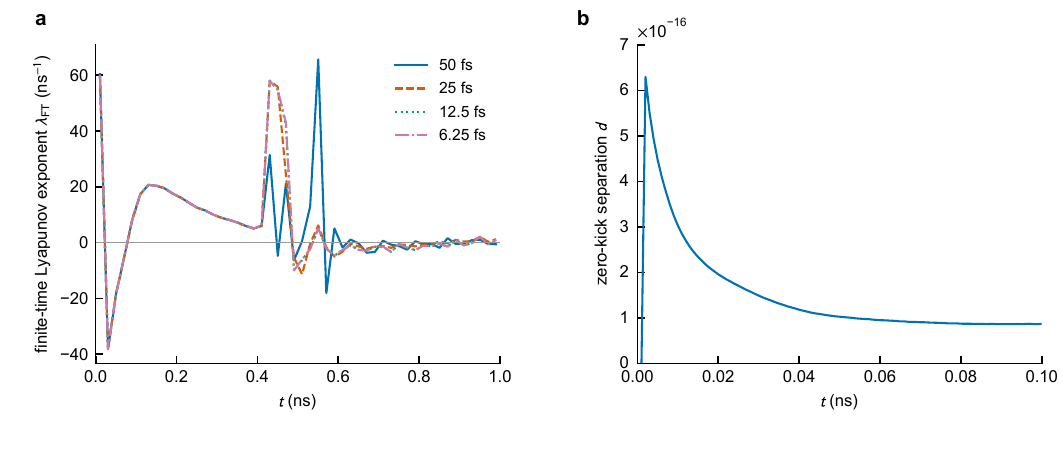}
    \caption{\textbf{\Rev{Numerical controls for finite-time Lyapunov exponents}.} \textbf{a}, \Rev{Finite-time} \Rev{Lyapunov} \Rev{exponents at fixed steps \mbox{$50$}, \mbox{$25$}, \mbox{$12.5$} and \mbox{$6.25$}~fs} (\Rev{blue} \Rev{solid,} \Rev{orange} \Rev{dashed,} \Rev{green} \Rev{dotted} \Rev{and} \Rev{purple} \Rev{dash-dotted}), \Rev{with} \Rev{\mbox{$\delta_0=10^{-6}$}} and \Rev{\mbox{$1$}}~ps \Rev{renormalization}. \textbf{b}, \Rev{Zero-kick} \Rev{RMS} \Rev{separation} \Rev{at \mbox{$6.25$}~fs, resetting} the \Rev{perturbed} \Rev{state} \Rev{to} \Rev{the} \Rev{reference} \Rev{after each interval. The \mbox{$12.5/6.25$}~fs difference} is \Rev{\mbox{$0.90\%$} in} the \Rev{specified} \Rev{active-window} \Rev{norm,} and the \Rev{measured} \Rev{zero-kick} \Rev{separation} \Rev{stays} \Rev{below} \Rev{\mbox{$6.3\times10^{-16}$}}.}
    \label{fig:lyapunov_si}
\end{figure*}

\begin{figure*}[!ht]
    \centering
    \includegraphics[width=\textwidth]{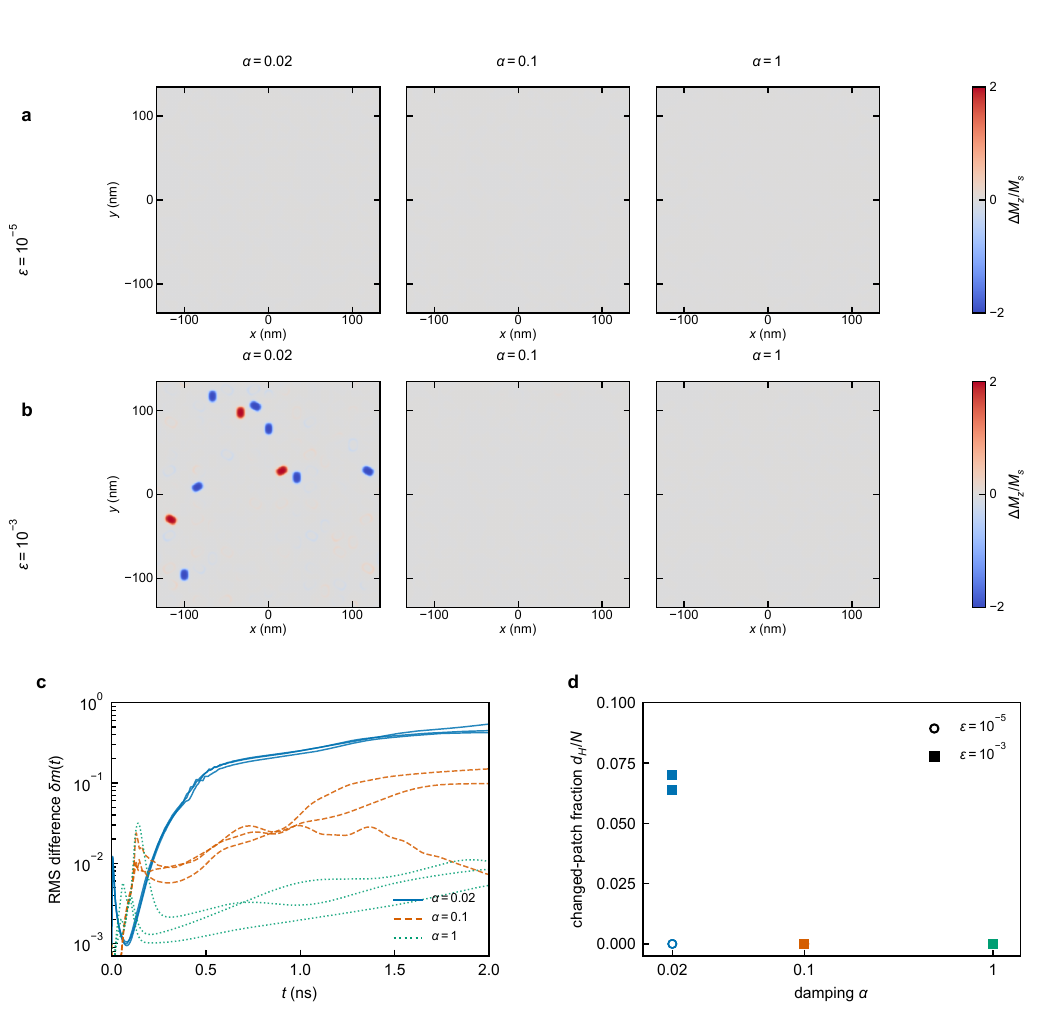}
    \caption{\textbf{\Rev{Damping dependence of transient amplification and domain sensitivity.}} \textbf{\Rev{a,b}}\Rev{, Signed top-layer differences \mbox{$\Delta M_z/M_s$} at \mbox{$2$}~ns for \mbox{$\varepsilon=10^{-5}$} and \mbox{$10^{-3}$}, respectively, using the first initial state. All maps share the main-text Fig.~\ref{fig:lyapunov} colour range.} \textbf{\Rev{c}}\Rev{, Unrescaled RMS separation \mbox{$\delta m(t)$} at \mbox{$\varepsilon=10^{-3}$}; each damping value has three traces, one per initial state.} \textbf{\Rev{d}}\Rev{, Final changed-patch fraction \mbox{$d_H/N$} for all three states, with \mbox{$N=157$}; open circles and filled squares denote the two amplitudes. Points may overlap. At \mbox{$\varepsilon=10^{-3}$}, \mbox{$10$}--\mbox{$11$} patches change for \mbox{$\alpha=0.02$}, versus none for \mbox{$0.1$} or \mbox{$1$}. Separation growth alone is not a local Lyapunov-rate measurement.}}
    \label{fig:damping_sensitivity}
\end{figure*}

\begin{figure*}[!ht]
    \centering
    \includegraphics[width=\textwidth]{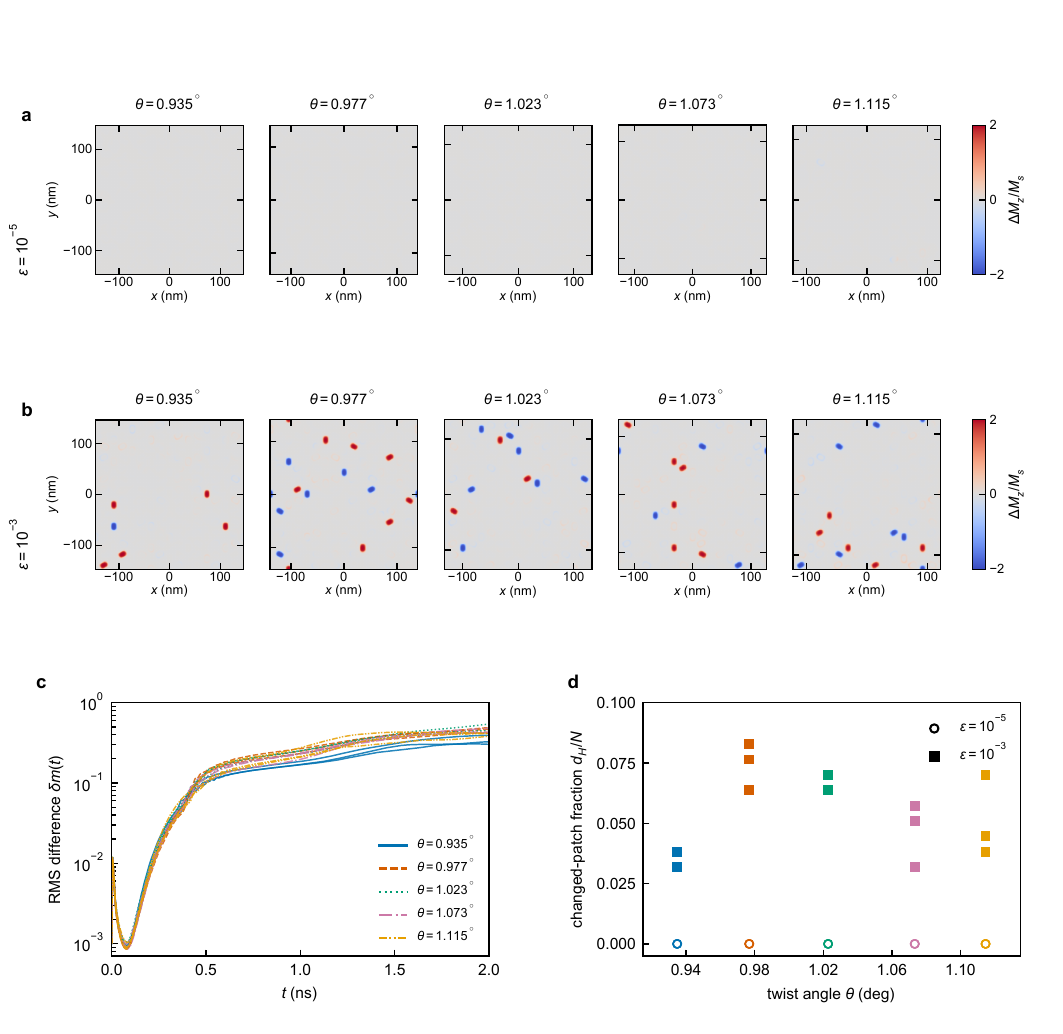}
    \caption{\textbf{\Rev{Twist-angle dependence of transient amplification and domain sensitivity.}} \textbf{\Rev{a,b}}\Rev{, Signed top-layer differences at \mbox{$2$}~ns} for \Rev{\mbox{$\varepsilon=10^{-5}$}} \Rev{and \mbox{$10^{-3}$}}, \Rev{respectively,} \Rev{using the first initial state at each angle} and the \Rev{same} \Rev{colour} \Rev{range} \Rev{as} \Rev{main-text} \Rev{Fig.~\ref{fig:lyapunov}.} \textbf{\Rev{c}}\Rev{,} \Rev{Unrescaled} \Rev{RMS} \Rev{separation} \Rev{at \mbox{$\varepsilon=10^{-3}$}, with three initial-state traces per angle}. \textbf{\Rev{d}}\Rev{, Final changed-patch fraction for both amplitudes (open circles and filled squares), showing all three states; points may overlap. At \mbox{$\varepsilon=10^{-3}$}, each angle shows \mbox{$5$}--\mbox{$13$} changed patches among \mbox{$157$}; none change at \mbox{$10^{-5}$}. All calculations use \mbox{$\alpha=0.02$} and \mbox{$T=0$}~K; these finite-amplitude comparisons do not measure a local Lyapunov exponent.}}
    \label{fig:twist_sensitivity}
\end{figure*}

\Rev{The fixed-step ladder is \mbox{$50$}, \mbox{$25$}, \mbox{$12.5$} and \mbox{$6.25$}~fs (Fig.~\ref{fig:lyapunov_si}a). The checked interval is \mbox{$0.05$}--\mbox{$0.5$}~ns, restricted to \mbox{$20$}~ps blocks whose absolute rate in the original \mbox{$25$}~fs trace exceeds \mbox{$1$}~ns\mbox{$^{-1}$}. The adjacent \mbox{$12.5/6.25$}~fs traces have relative RMS difference \mbox{$0.00905$}, below the specified \mbox{$0.05$} criterion. At \mbox{$6.25$}~fs, the reset amplitudes \mbox{$10^{-8}$} and \mbox{$10^{-6}$} differ by \mbox{$1.11\times10^{-4}$} in the same norm. The coarse-step traces are retained in the figure; the passing comparison applies to the stated interval and initial state.}

\Rev{A zero-kick control resets the perturbed state to the reference after every \mbox{$1$}~ps interval and runs for \mbox{$0.1$}~ns. Its maximum measured separation is \mbox{$6.3\times10^{-16}$} (Fig.~\ref{fig:lyapunov_si}b), far below the nonzero reset amplitudes. The perturbed calculations cover \mbox{$1$}~ns. For \mbox{$\delta_0=10^{-6}$}, the finite-time} Lyapunov exponent \Rev{averaged over \mbox{$0.05$}--\mbox{$0.5$}~ns is \mbox{$15.24$}~ns\mbox{$^{-1}$}, whereas the average over \mbox{$0.5$}--\mbox{$1$}~ns is \mbox{$-1.31$}~ns\mbox{$^{-1}$}. The sign change of the finite-time Lyapunov exponent resolves} transient \Rev{expansion} \Rev{followed by contraction during relaxation. The finite-amplitude amplification rate in main-text Fig.~\ref{fig:lyapunov}d is defined by Eq.~\eqref{eq:lyapunov}.}

\clearpage
\section{\Rev{Damping dependence of transient amplification.}}\label{sec:damping_sensitivity}

\Rev{We compare \mbox{$\alpha=0.02$}, \mbox{$0.1$} and \mbox{$1$} at \mbox{$\theta=1.023^\circ$}, using the material constants of main-text Figs.~\ref{fig:lyapunov}--\ref{fig:thermal_locking}. Three initial states (seeds \mbox{$20260824$}--\mbox{$20260826$}) each use one fixed tangent direction (seeds \mbox{$911000$}--\mbox{$913000$}) at \mbox{$\varepsilon=10^{-5}$} and \mbox{$10^{-3}$}. Single-precision adaptive trajectories run for \mbox{$2$}~ns at \mbox{$T=0$}~K, with tolerance \mbox{$10^{-6}$} and \mbox{$2$}~ps output. For the first state, tolerance \mbox{$10^{-7}$} preserves the final patch counts; all zero-kick controls remain identical.} The \Rev{\mbox{$\alpha=1$}} \Rev{case} \Rev{is a strong-damping control.}

\clearpage
\section{\Rev{Twist-angle dependence of transient amplification.}}\label{sec:twist_sensitivity}

\Rev{At \mbox{$\alpha=0.02$}, we use \mbox{$\theta=0.935^\circ$}, \mbox{$0.977^\circ$}, \mbox{$1.023^\circ$}, \mbox{$1.073^\circ$} and \mbox{$1.115^\circ$}. The grids are \mbox{$644\times651$}, \mbox{$616\times623$}, \mbox{$588\times595$}, \mbox{$560\times567$} and \mbox{$540\times546$}, respectively, with unchanged cell dimensions and material constants. Each angle uses a periodic cubic-spline resampling} of the \Rev{native} \Rev{exchange} \Rev{map} \Rev{followed} by \Rev{\mbox{$4\times7$}} \Rev{tiling}, an \Rev{approximation} rather than a \Rev{new} \Rev{first-principles} \Rev{calculation. The initialization recipe, perturbations and solver controls are as above;} the \Rev{different} \Rev{grids do not share identical spin arrays. All angles retain \mbox{$157$} interior patches. The refined tolerance preserves the first state's final patch counts.}

\end{document}